\documentclass[sn-mathphys,oneside]{sn-jnl}

\usepackage{manyfoot}  
\usepackage{xcolor}    
\usepackage{appendix}
\usepackage{siunitx} 
\usepackage{booktabs} 
\usepackage{makecell} 
\usepackage{amsmath,amssymb,amsfonts}  
\usepackage{bm}                        
\usepackage{graphicx}
\usepackage{hyperref}
\usepackage{svg}
\usepackage{amsmath}
\usepackage{tabularx}
\usepackage{lipsum}
\usepackage[section]{placeins}
\newcommand{\bs}[1]{\bm{#1}}

\usepackage[english]{babel}
\usepackage[most]{tcolorbox}
\usepackage[utf8]{inputenc}

\usepackage{textgreek}
\usepackage{soul} 
\usepackage{tcolorbox} 
\usepackage{xcolor} 
\usepackage[normalem]{ulem}

\newcommand{\rank}{\hbox{\rm rank}}

\definecolor{LightGray}{gray}{0.9}

\setcitestyle{numbers,square,citesep={,}}

\usepackage{listings}
\usepackage{chngcntr}
\usepackage{hyperref} 

\counterwithin{table}{section}

\begin{document}

\title{ALL--FEM: Agentic Large Language models Fine-tuned for Finite Element Methods}

\author[1]{\fnm{Rushikesh} \sur{Deotale}}\email{rdeotal@purdue.edu}
\author[1]{\fnm{Adithya} \sur{Srinivasan}}\email{srini162@purdue.edu}
\author[1]{\fnm{Mahmoud} \sur{Golestanian}}\email{mgolesta@purdue.edu}
\author[2]{\fnm{Yuan} \sur{Tian}}\email{tian211@purdue.edu}
\author[2]{\fnm{Tianyi} \sur{Zhang}}\email{tianyi@purdue.edu}
\author[1]{\fnm{Pavlos} \sur{Vlachos}}\email{pvlachos@purdue.edu}
\author*[1]{\fnm{Hector} \sur{Gomez}}\email{hectorgomez@purdue.edu}

\affil[1]{\orgdiv{School of Mechanical Engineering}, \orgname{Purdue University}, \orgaddress{\street{585 Purdue Mall}, \city{West Lafayette}, \postcode{IN 47907}, \country{USA}}}
\affil[2]{\orgdiv{Department of Computer Science}, \orgname{Purdue University}, \orgaddress{\street{305 N University St}, \city{West Lafayette}, \postcode{IN 47907}, \country{USA}}}

\abstract{
Finite element (FE) analysis guides the design and verification of nearly all manufactured objects. It is at the core of computational engineering, enabling simulation of complex physical systems, from fluids and solids to multiphysics systems. However, implementing FE codes and analyzing simulation results demands expertise across numerical analysis, continuum mechanics, and programming. Conventional Large Language Models (LLMs) can generate FE code, but they hallucinate, lack awareness of variational structures, and cannot close the loop from problem statement to a verified solution. Here, we propose ALL-FEM, an autonomous simulation system that integrates agentic AI with domain-specific, fine-tuned LLMs for FEniCS code generation across solid, fluid, and multiphysics applications. We construct a corpus of 1000+ verified FEniCS scripts by combining 500+ curated expert codes with a retrieval-augmented, multi-LLM pipeline that generates and filters codes for diverse PDEs, geometries, and boundary conditions. We used the corpus to fine-tune LLMs with 3B to 120B parameters. Our agentic framework orchestrates specialized agents, powered by fine-tuned LLMs, to formulate problems as PDEs, generate and debug code and visualize the results. We evaluated the system on 39 benchmarks that include problems of linear/nonlinear elasticity, plasticity, Newtonian/non-Newtonian flow, thermofluids, fluid-structure interaction, phase separation, and transport on moving domains. Embedded in a multi-agent workflow with runtime feedback, the best fine-tuned model (GPT OSS 120B) achieves code-level success of 71.79\%, outperforming a non-agentic deployment of GPT 5 Thinking. By showing that relatively small, fine-tuned LLMs, orchestrated through agentic frameworks, can automate FE workflows, ALL-FEM offers a blueprint for autonomous simulation systems in computational science and engineering.

}
\maketitle


\section{Introduction}

Finite element analysis is the cornerstone of computational engineering, enabling computer simulation of the behavior and response of complex physical systems \cite{hughes2012finite, logg_automated_2012, zienkiewicz2005finite, HUGHES20054135}. Examples include fluids, solids, combinations thereof and many other systems controlled by multiphysics processes. Nearly all manufactured objects, from aircrafts and bridges to biomedical devices, are designed using information from finite element analysis. Despite the prevalence of finite elements, implementing the algorithm and deploying it remain time-consuming, repetitive and high-competence tasks. Even with high-level libraries such as FEniCS \cite{alnaes_fenics_2015, logg_automated_2012} or Firedrake \cite{FiredrakeUserManual}, writing, testing, and validating efficient codes requires knowledge of variational formulations, time integration and linear algebra. In academic research, in-house and open software prevail. As a result, countless hours are spent developing incrementally similar code, e.g., modifying a code for compressible elasticity to incompressible elasticity. This results in duplicated efforts with minimal added value. In industry, commercial software dominates because it provides a sense of reliability through black-box operation, minimizing user-induced numerical or formulation errors. However, this comes at high licensing and operational costs, and it inherently limits the ability to solve non-standard problems.

The emergence of Large Language Models (LLMs) has opened new possibilities for automating computational mechanics workflows that can simultaneously address the inefficiencies in the research and industry ecosystems \cite{guo2026large,ni2023mechagentslargelanguagemodel}. However, off-the-shelf LLMs often generate code with hallucinations \cite{zhang2025sirenssongaiocean, chang2023surveyevaluationlargelanguage, agarwal2025codemiragehallucinationscodegenerated} including syntactic or semantic errors, leading to incorrect results. This unreliability limits its adoption in contexts where correctness is critical. To mitigate hallucinations in LLMs, one strategy is to fine-tune them on domain-specific datasets. LLMs are typically pre-trained on vast amounts of data from the internet, and fine-tuning helps them internalize the specific syntactic patterns of FEM codes, constraining their responses to valid formulations that embed knowledge of variational formulations and finite element technology. A complementary approach to increase reliability is to replace a zero-shot LLM that generates a single output per prompt with agentic AI systems, that can reason sequentially, verify intermediate steps, and perform actions by interacting with external tools. 

Recent research \cite{li2024agentsneed} has shown that multi-agent AI systems in which each agent has a specific role can improve reliability in problem formulation, tool use, and simulation execution. Examples of successful agentic systems include applications in elasticity \cite{ni2023mechagentslargelanguagemodel,tian2024optimizingcollaborationllmbased}, protein discovery \cite{Ghafarollahi2024} and materials discovery \cite{zhang2024honeycombflexiblellmbasedagent}. Many of the efforts in computational mechanics have focused on computational fluid dynamics (CFD). OpenFOAMGPT used retrieval-augmented generation (RAG), a method that uses an information retrieval step before generating an output \cite{lewis2020rag}, to solve problems of turbulent flows, heat transfer and multiphase flows \cite{pandey2025openfoamgpt,wang-2025}. OpenFOAMGPT 2.0 \cite{feng2025openfoamgpt2} extended this framework to a multi-agent architecture improving the results of OpenFOAMGPT. Foam-Agent explored hierarchical multi-index RAG with dependency-aware generation to improve configuration consistency across the CFD workflow \cite{yue2025foamagent}. AutoCFD aimed at enabling direct translation from natural language to executable CFD setups by fine-tuning on a large database of language-to-OpenFOAM configuration pairs with chain of thought annotations \cite{dong2025autocfd}. Similar efforts include MetaOpenFOAM, which integrates RAG-enhanced knowledge retrieval with an agentic framework to decompose complex CFD tasks \cite{chen2024metaopenfoam} and ChatCFD, which uses iterative trial-reflection-refinement loops to address simulation cases from the literature \cite{fan2025chatcfd}. New efforts are emerging beyond CFD. AutoFEA combines a graph convolutional network (GCN)-transformer model with LLMs for automating FE code generation \cite{hou2025autofea}. FeaGPT develops a framework for structural FE analysis that attempts to automate the complete workflow from geometry generation to simulation results \cite{feagpt}. There is also interest in automating scientific discovery beyond computational workflows. Reference \cite{feng2025turbulenceai} introduced turbulence.ai, an AI scientist for fluid mechanics with capabilities to generate hypotheses, perform simulations and draft a manuscript. AGENTICSCIML is a multi-agent system for discovery in scientific machine learning. The agents cooperate to propose, critique, and refine scientific machine learning solutions through structured reasoning and iterative evolution \cite{agenticsciml}. ATHENA is an agentic framework that operates as an autonomous lab that manages the complete research lifecycle \cite{athena}. Guo et al.~\cite{guo2026large} proposed an LLM-empowered CAE agent, where LLMs act as autonomous collaborators that plan, execute, and adapt CAE workflows, and apply it to data-free intrusive model order reduction to reduce solver redevelopment effort~\cite{benner2015projection_mor_survey}. Using Tensor-decomposition-based A Priori Surrogates (TAPS)~\cite{guo2025taps_arxiv}, they show that natural-language descriptions of parametric PDEs can be translated into solver implementations that yield reduced-order models.

Here, we develop ALL-FEM, an autonomous simulation system that uses agentic AI and fine-tuned LLMs to automate substantial portions of the computational mechanics workflow. Our framework includes domain-specific, fine-tuned LLMs for FEniCS code generation. We fine-tuned open-weight LLMs ranging in size from 3B to 120B on a FEniCS code dataset with over 1000 entries. To address data scarcity, we propose a data augmentation pipeline in which multiple LLMs collaborate sequentially to expand a small, expert-curated seed dataset into a much larger, diverse database of high-quality, correct FEniCS codes. ALL-FEM introduces a novel end-to-end framework that (i) creates a verified synthetic dataset tailored to mapping boundary value problems for PDEs to FEniCS code; the synthetic dataset pipeline provides a new approach for generating verified code datasets in domains with limited publicly available data, with applications potentially extending to other software domains with data constraints similar to FEniCS code; (ii) uses that dataset to fine-tune an open source LLM that generates executable and reliable FEniCS codes for FEM problems, and (iii) deploys this fine-tuned FEM-specialized model within an automated multi-agent workflow. Within this workflow, the system accomplishes multiple critical steps in the simulation, including expressing the problem in weak form with an adequate formulation of the boundary conditions; generating an initial FEniCS implementation of the problem; debugging the code; checking consistency between formulation and implementation; and plotting the numerical results. The ALL-FEM framework links data generation, model specialization, and automation of the simulation workflow in one end-to-end pipeline. An important feature of our agentic system is that the selection and coordination of agents is orchestrated by an LLM, enabling flexibility, adaptability and robustness across different computational mechanics challenges.

We evaluated our agentic systems using a benchmark composed of 39 problems that include linear/nonlinear elasticity, plasticity, Newtonian/non-Newtonian flow, thermofluids, fluid-structure
interaction, simplified multiphase flow models, and diffusion-reaction systems on moving domains. When used in a multi-agent framework, our best fine-tuned LLM produces correct code in 71.79\% of the benchmark problems, and outperforms a non-agentic deployment of the OpenAI model GPT-5 Thinking. Our results show that even relatively small models, fine-tuned with a modest database, can reliably produce correct finite element codes when embedded within an appropriate agentic framework. Our agentic system can simultaneously advance computational science across the research enterprise and industry and opens new opportunities for autonomous simulation systems, in which the entire finite element workflow, from problem statement to validated solution, is automated by coordinated, specialized LLM agents.

\section{Methods}

\subsection{LLM Selection}
In recent years, LLMs built on the transformer architecture have progressed rapidly and become useful general tools \cite{chang2023surveyevaluationlargelanguage, anil2023palm2technicalreport, https://doi.org/10.48550/arxiv.1706.03762}. 
Contemporary models are specialized for diverse tasks. Long-context variants now support very large inputs, enabling extended problem descriptions \cite{anthropic2023100k}. Furthermore, LLMs exhibit proficiency in code synthesis; code-tuned models such as Codex demonstrate functional program generation on HumanEval, a dataset of human-created problems to benchmark LLMs on code generation \cite{chen2021evaluatinglargelanguagemodels}. Recent work has shown that reasoning can be elicited with chain-of-thought prompting and further improved by self-consistency decoding \cite{wei2023chainofthoughtpromptingelicitsreasoning, wang2023selfconsistency}. Due to these capabilities,  LLMs have the potential to assist or partially replace human effort in problem setup, coding, debugging, and post-processing for computational mechanics.

One element of the work presented herein requires fine-tuned LLMs specialized in writing finite element codes using FEniCS. To develop these fine-tuned LLMs, we identify models that (1) are open-source and allow fine-tuning, and (2) have proven capabilities in code generation and logical reasoning. Following an evaluation of various open-source LLMs on Hugging Face, we selected the following models for fine-tuning: Llama 3.2 3B \cite{meta2024llama32}, Qwen3 32B \cite{qwen3_32b_hf_2025}, Llama 3.3 70B \cite{meta_llama33_70b_instruct_2024}, and GPT-OSS 120B \cite{openai2025gptoss120b}. For all models, the suffix (e.g., 3B, 32B, 70B, 120B) indicates the number of trainable parameters. This selection of models spans a wide range of scales and architectures, enabling us to study how these factors affect performance. For comparison, we will benchmark the performance of our fine-tuned models against the commercial model, GPT-5 Thinking, from OpenAI \cite{openai2025_gpt5_system_card}. Because GPT-5 Thinking is proprietary, it cannot be fine-tuned. However, it serves as a state-of-the-art reference baseline. Below, we briefly describe the capabilities of the selected models and explain the reasons behind our choices. 

\begin{enumerate}  

\item Llama 3.2 3B was developed by Meta AI \cite{meta2024llama32}. For its relatively small size, Llama 3.2 3B performs very well at reasoning, achieving a notable 77.7\% in the math reasoning section of GSM8K \cite{meta_ai_2024_llama32}.

\item Qwen 3 32B was developed by Alibaba Cloud. This model has proven good performance in coding. Public benchmarks, such as the Aider Polyglot suite, which analyze code-editing capabilities, show that Qwen3 32B outperforms all comparable models of its size   \cite{aider_llm_leaderboards_2025}. 

\item Llama 3.3 70B exhibits strong performance on tasks that require reasoning and coding abilities. Meta's technical research indicates that Llama 3.3 70B exhibits high performance on numerous coding tasks \cite{llama3_herd_2024}. Independent evaluations by Vellum AI showed that Llama 3.3 70B performs on par with larger models like GPT-4 on tasks of medium complexity, at a lower operational cost \cite{vellum_llama33_vs_gpt4o_2024}.  
  
\item GPT-OSS 120B, developed by OpenAI, is one of the most advanced open-source models available for fine-tuning \cite{openai2025_introducing_gpt_oss}. This model uses a Mixture-of-Experts (MoE) architecture, which activates around 5.1B parameters per token during inference \cite{openai2025_gpt_oss_github}. This approach allows the model to handle very long inputs (128K tokens), making it well-suited to situations that involve extensive reasoning and vast code contexts \cite{openai2025_introducing_gpt_oss}. GPT-OSS 120B achieves performance similar to that of OpenAI’s proprietary o4-mini model on benchmarks for general problem-solving (MMLU \cite{hendrycks2021measuringmassivemultitasklanguage} and GPQA Diamond \cite{rein2023gpqagraduatelevelgoogleproofqa}), matches the performance of OpenAI o3 in mathematics competitions (AIME 2024/2025 \cite{patel2024aimeaioptimizationmultiple}) \cite{openai2025_gpt_oss_model_card} and outperforms OpenAI o1 and GPT-4o at HealthBench--- a benchmark for LLM performance in healthcare \cite{openai2025_introducing_gpt_oss}. Because the model uses MXFP4 quantization for the MoE weights, it can be deployed on a single high-end GPU such as the NVIDIA H100 \cite{openai2025_gpt_oss_github, hf_transformers_mxfp4}. The inclusion of GPT-OSS 120B in our study provides a key benchmark for understanding how modern, generation-efficient models that use MoE compare with older dense models that use standard inference.  

\item GPT-5 is the current flagship model from OpenAI \cite{openai2025_gpt5_system_card} and is one of the strongest publicly available general-purpose LLMs. The Thinking version of GPT-5 used here is particularly strong at coding and reasoning, achieving higher scores than GPT-OSS 120B and earlier OpenAI models like OpenAI o3 on AIME 2025 and on coding benchmarks such as SWE-bench Verified \cite{openai2025introducinggpt5,patel2024aimeaioptimizationmultiple,openai2025_gpt_oss_model_card}. Although OpenAI has not disclosed the exact parameter count, estimates indicate that GPT-5 has well over 1 trillion parameters, making it by far the largest model in our study.


\end{enumerate}

We will fine-tune all the models listed above, except GPT-5. To refer to the fine-tuned version of a model, we will add FT at the end of its original name. For example, we will call Llama 3.2 3B FT our fine-tuned version of Llama 3.2 3B. In what follows, we describe our fine-tuning strategy.

\subsection{Fine-tuning}

Off-the-shelf LLMs are trained on large-scale datasets built from internet text such as websites, books, scientific articles and other publicly available corpora.  They excel at general knowledge and problem-solving, but they struggle to generate reliable legacy FEniCS code for computational mechanics tasks.
Fine-tuning pre-trained models on domain-specific datasets improves their performance on domain-specific tasks. This has been shown in studies such as BatteryBERT for battery research \cite{Huang2022}, BioBERT for biology \cite{Lee_2019}, and ProGen for protein engineering \cite{Madani2023}.   
Fine-tuning an LLM involves further training a pre-trained model on a smaller, domain-specific dataset to shift its parameters toward the target domain, thereby improving its performance \cite{han2024parameterefficientfinetuninglargemodels}.
In this study, we fine-tune our selected open-source LLMs to improve their ability to generate reliable computational mechanics code using FEniCS. This is part of a separate, ongoing effort by our research group to develop {\tt FEniCS-LLM}, an open-source LLM that outperforms existing LLMs in FEniCS code generation. Both the model and the dataset used for fine-tuning are open and available at \href{https://fenics-llm.github.io}{https://fenics-llm.github.io}.


\subsubsection{Dataset Creation}

To fine-tune our models, we first create a dataset containing computational mechanics problems and the FEniCS codes that solve them. The dataset is structured in the Alpaca format, where each data entry contains an instruction, input, and output triple \cite{taori_alpaca_2023}. The instruction provides a high-level description of the problem, for example: ``Develop a Python script using the FEniCS library to solve the 2D Poisson equation on a rectangular domain with Dirichlet boundary conditions''. The input describes more details of the problem, such as the governing equations, domain geometry, and the specific values of the boundary conditions. The output provides a ground-truth solution in the form of an executable FEniCS script that correctly solves the problem described in the instruction and input.

The dataset was created in two stages. In the first stage, we curated a seed dataset, collecting FEniCS codes from established and reputable sources on the internet as well as our in-house codes. As the resources on the internet and our research group were limited, we developed an automated pipeline to augment the data. In the second stage, the seed dataset was used as input to the automated pipeline that generated additional synthetic samples. The dataset after data augmentation consists of 1004 FEniCS codes that solve PDEs specific to the domain of computational mechanics.

\paragraph{Seed Data Collection}
The seed data was collected from the FEniCS tutorials \cite{langtangen_fenics_2017}, lecture notes for MAE 207: Finite Element Analysis for Coupled Problems by David Kamensky \cite{Kamensky_MAE207_FEA}, legacy FEniCS demos from the official DOLFIN documentation \cite{Fenics_Dolfin_Demos}, instructional materials from subject-matter experts, and publicly accessible code archives maintained by the University of Florida and the University of South Carolina \cite{Burkardt_FEniCS_Examples}. In addition, we incorporated approximately 40 in-house scripts developed by our research group. This data curation yielded 503 entries, which served as the seed database for subsequent synthetic data generation. The seed dataset contains approximately 95 solid mechanics entries, covering linear elasticity, hyperelasticity, and plasticity. Fluid mechanics is represented by around 90 entries, encompassing Stokes flow, Navier-Stokes flow, Darcy flow, and the Burgers' equation. We also included 50 multiphysics entries for coupled problems, such as fluid-structure interaction, phase-field models, and thermoelasticity. Additionally, 70 entries focus on heat transfer and other diffusion equations. The remaining 200 entries cover mathematical fundamentals and core software functionality. Instead of representing a specific physical phenomenon, these scripts solve standard PDEs (e.g., Poisson's equation) or demonstrate essential coding tasks, such as generating meshes, defining boundary conditions, and configuring linear solvers.

\begin{figure} 
  \centering
  \includegraphics[width=\textwidth,keepaspectratio]{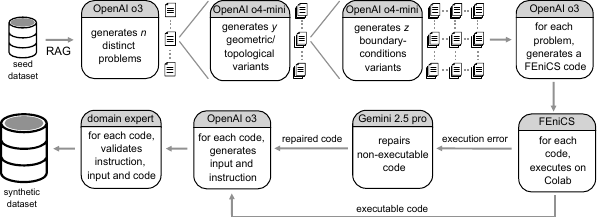}
  \caption{Flowchart describing our data augmentation technique. We use a multi-model pipeline to create synthetic FEniCS codes from a seed dataset.}
  \label{fig:dataset}
\end{figure}

\paragraph{Synthetic Data Generation Pipeline}
Fig.~\ref{fig:dataset} shows our data augmentation pipeline. The pipeline is based on a multi-model workflow and leverages retrieval-augmented generation (RAG)--- a method that uses an information retrieval step before generating an output \cite{lewis2020rag}. {In the first step of the pipeline, we used the entire seed dataset as the knowledge base for RAG and used OpenAI o3 to generate new, unique problem statements covering $n=7$ distinct problems, each with its own PDE.} These include the Poisson equation for electrostatics, the heat conduction equation, linear elasticity, the Stokes equations for fluid flow, the advection-diffusion equation, the Helmholtz equation, and plasticity problems. Each of the $n$ problems proposed by OpenAI o3 was then passed to an OpenAI o4 mini model, which generated $y=10$ copies of the same problem across different computational domains. The geometric variants included domains defined by squares, rectangles, triangles, circular disks, elliptical disks, trapezoids and L-shaped polygons. The topological variants included different homotopy types created by taking one geometric variant and making a hole in its interior with the shape of another geometric variant. Some possible outcomes of this process are annuli, or L-shaped domains with a circular hole. 
{Following this, each geometric or topological variant was sent to OpenAI o4-mini to generate $z=10$ copies of each problem with different boundary conditions.} {To balance data quality and computational cost, we used the more capable OpenAI o3 model for the conceptually demanding steps, such as designing diverse, well-posed PDE problems. The lighter, more cost-efficient OpenAI o4-mini handles systematic geometry and boundary-condition variants, as it maintains strong coding performance at lower cost than larger models} \cite{openai2025_o3_o4mini}. For each resulting problem variant, OpenAI o3 was prompted to create a FEniCS code. In total, we generated $nyz=700$ unique FEniCS codes. Every FEniCS code was then tested for execution on Google Colab \cite{google-colab}. We used Gemini 2.5 pro to correct the codes that resulted in an execution error via one-shot prompting. {We chose Gemini 2.5 Pro for this stage because it combines strong code-understanding and debugging capabilities} \cite{google2025_gemini25_blog} {integrated with the Google Colab environment, allowing us to diagnose errors and fix code directly within the Colab notebook}. The codes that resulted in error after one correction attempt were discarded. The results produced by the executable codes were verified for correctness by domain experts in our group. Because each synthetically created instance is unique, we could not obtain reference solutions (e.g., analytical or benchmark findings) for every case; instead, we utilized a structured expert evaluation. Human reviewers first examined prompt-to-code compatibility (governing equation, weak form, spaces, geometry, mesh, and boundary conditions) and eliminated any executable but incorrect implementations. Then, they reviewed outputs in ParaView to validate the geometry, boundary values, and check if qualitative solution behavior matched expectations. After a FEniCS code was successfully evaluated, we used OpenAI o3 to generate the corresponding instruction and input fields, completing the instruction-input-output record for each script. Through this pipeline, we obtained 501 additional dataset entries, thereby substantially extending our dataset without compromising data quality.

\subsubsection{Low-Rank Adaptation (LoRA) and Quantized LoRA (QLoRA)}
Fine-tuning large pretrained models typically requires updating all parameters, which is computationally demanding and often unnecessary for downstream tasks. In practice, the parameter updates learned during task adaptation tend to concentrate in a much smaller subspace than the full weight space, suggesting that the effective degrees of freedom of the update are low~\cite{hu2021loralowrankadaptationlarge}. The LoRA approach to fine-tuning formalizes this observation by modeling the weight update as a low-rank matrix that is added to the pretrained weights. This is accomplished by keeping the original weights fixed while learning a small set of additional parameters, which are inserted in parallel to the existing model layers. The additional LoRA parameters and original weights will collectively determine the generation during inference. By avoiding updates to the entire model, LoRA dramatically reduces the number of trainable parameters, thereby lowering memory usage, shortening training time, and minimizing the storage cost of task-specific adaptations. Despite its lightweight form, LoRA typically reaches performance comparable to full fine-tuning.

Let $\bs{W}_{0}$ be a matrix that contains the weights of a pretrained model. During fine-tuning, we will determine a weight update matrix $\Delta\!\bs{W}$ such that the updated weights are $\bs{W} = \bs{W}_{0} + \Delta\!\bs{W}$. LoRA approximates $\Delta\!\bs{W}$ via a low-rank factorization. For an update $\Delta\!\bs{W}\in \mathbb{R}^{d \times k}$, we introduce two smaller matrices, $\bs{B} \in \mathbb{R}^{d \times r}$ and $\bs{A} \in \mathbb{R}^{r \times k}$, where $r=\rank(\bs{A})=\rank(\bs{B})$ verifies  $r \ll \min(d,k)$. The update is then written as the product $\Delta\!\bs{W} = \bs{B}\,\bs{A}$. The number of trainable parameters (number of entries in $\bs{A}$ and $\bs{B}$) is $d r + r k$, which is much smaller than $d k$ (number of entries in $\bs{W}_{0}$) when $r$ is small. By choosing a smaller $r$, we trade off expressive power for a lower parameter count. In practice, the weight update $\Delta\!\bs{W}$ is not directly added to the original weights, but it is scaled by a hyper‑parameter $\alpha$ and the rank $r$, such that 
\[
\bs{W} \;=\; \bs{W}_{0} \;+\; \frac{\alpha}{r}\,\bs{B}\,\bs{A}.
\]
Here, $\alpha$ controls the overall influence of the learned LoRA update, while $r$ determines the LoRA module's capacity to learn new information. Both are selected before training based on practical needs. A larger $\alpha$ increases the impact of the update but may lead to forgetting original knowledge. In contrast, a larger $r$ improves learning capacity at the cost of greater memory and computational overhead. The matrices $\bs{A}$ and $\bs{B}$, together with their hyperparameters $(r,\alpha)$ and the choice of attention and feed-forward projection layers to which they are attached, is referred to as a LoRA adapter \cite{hu_lora_2022, dettmers2023qloraefficientfinetuningquantized}

The modified forward pass for the layer becomes

\[
\bs{h} = \bs{W}\,\bs{x} = 
\bs{W}_{0}\,\bs{x} \;+\; \frac{\alpha}{r}\,\bs{B}\,\bs{A}\,\bs{x},
\]
where $\bs{x}$ is the input and $\bs{h}$ the output. Because the update is folded into $\bs{W}$, LoRA introduces no additional inference latency compared with the original model. This feature makes LoRA a highly effective method for adapting LLMs to specialized domains without incurring the prohibitive costs of full fine‑tuning.

QLoRA (Quantized LoRA) is a memory-efficient variant of LoRA that quantizes (i.e., reduces the precision of weight storage, such as converting from float to integer) the pre-trained model weights, which are kept fixed during fine-tuning, down to 4-bit precision. Apart from this quantization of the base model weights, the fine-tuning procedure is identical to that used with standard LoRA \cite{dettmers2023qloraefficientfinetuningquantized}. Due to hardware memory constraints, the larger Llama 3.3 70B  and GPT-OSS 120B were fine-tuned using QLoRA, whereas we fine-tuned Llama 3.2 3B and Qwen 3 32B using standard LoRA.

\subsubsection{Hyperparameters and Hardware}

In this subsection, we (1) describe the settings of the LoRA and QLoRA adapters, and (2) specify the optimizer, learning-rate schedule, batch size, and number of optimization steps, (3) describe the hardware used for fine-tuning.

\paragraph{LoRA Settings}
In our experiments, each LoRA module is configured with rank $r=8$,  a common choice in LoRA fine-tuning. 
The scaling factor $\alpha$ determines the overall strength of the update through the ratio $\alpha/r$. We select $\alpha=2r$, this follows common practice in LoRA \cite{shuttleworth2024lora, kalajdzievski2023rankstabilizationscalingfactor, biderman2024lora}. To reduce over-fitting of the adapter parameters on our relatively small fine-tuning dataset, we also apply dropout inside each adapter. Dropout randomly sets a fraction of activations to zero during fine-tuning. We use a dropout rate of 
0.15 on the adapter activations. We attach LoRA adapters to all attention and feed-forward projection layers in the transformer, so that every linear projection in the transformer blocks is updated through a low-rank modification during fine-tuning.


\paragraph{Learning Optimizer Settings}
Training was conducted using the HuggingFace \texttt{SFTTrainer} framework. The trainable LoRA parameters were optimized using the AdamW algorithm~\cite{adamw}. The AdamW optimizer was configured with $\beta_{1}=0.9$ and $\beta_{2}=0.999$, representing the exponential decay rates for the first- and second-moment estimates of the gradient, respectively, and a numerical stability term $\varepsilon = 1 \times 10^{-8}$. We chose the learning rate as follows: To stabilize early training dynamics, we started the training with a warm-up period in which the learning rate increased linearly from 0 to $3 \times 10^{-5}$. After that, the learning rate decreased linearly. We used a micro-batch size of 8192 tokens with gradient accumulation over 16 steps, resulting in an effective batch size of 16 samples. The training process was terminated after 210 weight updating steps, corresponding to 3.3 epochs on average.

\paragraph{Hardware}
All base and fine-tuned LLM models are hosted on Purdue’s Anvil and accessed through AnvilGPT, an on-prem LLM service for ACCESS researchers \cite{rcac2025anvilgpt}. Within Anvil, LLMs are served through an Ollama backend, a lightweight inference runtime for open-weight models \cite{ollama2025}. AnvilGPT provides an OpenAI-style chat-completion API, allowing us to use AutoGen with these models \cite{rcac2025anvilgpt_api}. AnvilGPT also provides features to create custom models with custom system prompts and the facility to RAG LLMs with files. All LLMs run on NVIDIA H100 GPUs on Anvil. For fine-tuning and validation, we used NVIDIA H100 GPUs on Purdue’s Gautschi cluster.

After fine-tuning with LoRA/QLoRA, the learned low-rank adapters were merged into their respective base models to reconstruct full 16-bit floating-point checkpoints for standalone inference \cite{huggingface_peft_lora_merge_2024}. We then utilized llama.cpp to convert these merged 16-bit floating-point models into the GGUF format \cite{llamacpp_repo_2023, llamacpp_hf_to_gguf_guide_2024}. GGUF is the inference format expected by downstream runtimes such as Ollama, for efficient GPU deployment \cite{huggingface_gguf_format_2024, ollama_import_gguf_2025}. The smaller models, Llama 3.2 3B and Qwen 3 32B, were kept in 16-bit floating-point GGUF without further quantization \cite{dettmers2023qloraefficientfinetuningquantized}. For the larger models, 4-bit quantization was used to fit models for single-GPU execution: Llama 3.3 70B was quantized using the Q4\_K\_M scheme. GPT-OSS 120B was quantized using IQ4\_NL, as Q4\_K\_M and Q4\_K\_S marginally exceeded the VRAM limit for inference on a single NVIDIA H100 (80 GB) \cite{artefact2_gguf_quants_2024}. After quantization, all GGUF models were imported into Ollama and packaged with appropriate configuration files (Modelfiles) for deployment on Purdue's Anvil cluster via Ollama’s GGUF-based runtime \cite{ollama_import_gguf_2025, ollama_modelfile_reference_2025}.

\subsubsection{Validation}

To obtain a strong estimate of model performance, we employ a 5-fold cross-validation strategy throughout our experiments. We split the dataset into five equal-sized folds. The model is trained on four folds and validated on the fifth set. The process is repeated five times, with each fold serving once as a validation set. We use 5-fold cross validation because it provides a good balance between reliable evaluation results and manageable computational cost \cite{raschka2020modelevaluationmodelselection, bradshaw2023guide}. 

\begin{figure} 
    \includegraphics[width=1\columnwidth]{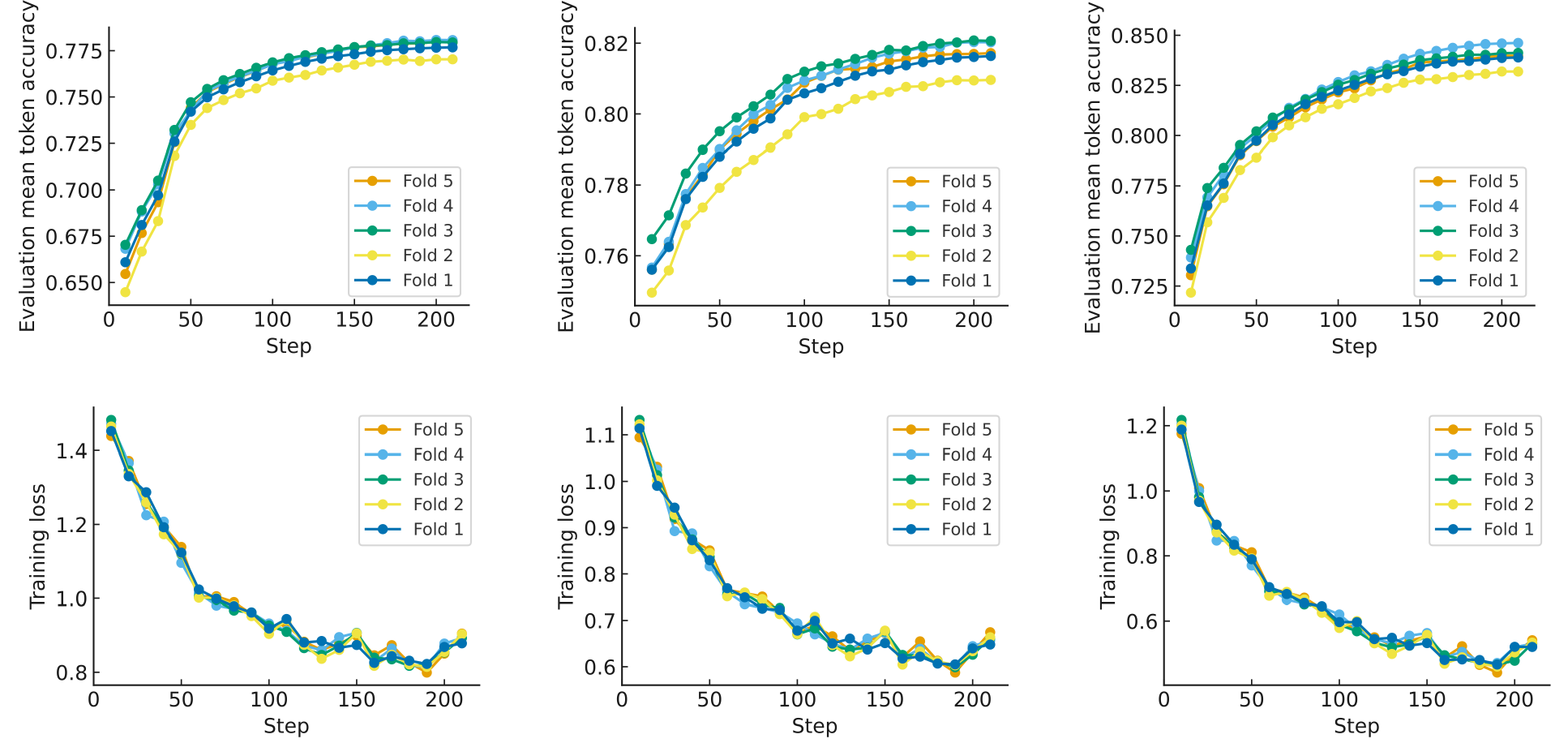}
    \caption{Evaluation mean token accuracy (top row) and training loss (bottom row) in a 5-fold cross-validation of our fine-tuning process. The left, center and right columns show results for Llama 3.2 3B, Qwen 3 32B, and Llama 3.3 70B, respectively.}
    \label{fig: validation}
\end{figure}
Fig.~\ref{fig: validation} presents the evaluation mean token accuracy over training steps (top row) and the corresponding training loss trajectory (bottom row). Left, center and right columns correspond, respectively, to LLaMA 3.2 3B, Qwen 3 32B and  LLaMA 3.3 70B. Due to computational constraints, we were unable to perform 5-fold cross-validation for GPT-OSS 120B. However, we discuss the effectiveness of fine-tuning GPT-OSS 120B in detail in the Results section. We fine-tune all models using supervised fine-tuning, i.e., training on labelled instruction-response pairs with a next-token prediction objective \cite{huggingface2025sfttrainer}. In this context, evaluation mean token accuracy is computed on the validation data as the proportion of token positions where the model’s highest-probability next token matches the reference token at that position, averaged over all validation examples \cite{huggingface2025sfttrainer}. A successful optimization run is expected to show an overall decrease in the training loss and an overall increase in the evaluation mean token accuracy across folds \cite{huggingface_llm_learning_curves, barinov2023automatic}. 

Our data show that the final mean token-level evaluation accuracy increases with the number of model parameters. These results align with existing scaling laws for LLMs, which indicate that larger models generally achieve lower cross-entropy loss when predicting the next token \cite{kaplan2020scalinglawsneurallanguage, yan2025scales}. Because the objective function in our supervised fine-tuning strategy is the cross-entropy, higher mean token accuracy usually leads to lower loss \cite{kaplan2020scalinglawsneurallanguage, yan2025scales}, which is also consistent with our data. In our fine-tuning process, the training loss was reduced by factors ranging from $\sim\!1.7$ (Llama 3.2 3B) to $\sim\!2.4$ (Llama 3.2 70B). The fastest decrease in the training loss occurs in the first $\sim\!50$ steps. Beyond step $\sim\!150$, all models exhibit flattened loss curves, indicating diminishing returns with further training. We can also see that in all three models there is very little change across folds. This points to a homogeneous and consistent dataset, with no single fold disproportionately affecting performance.

\subsection{Agentic AI}

Although our fine-tuned LLMs produce high-quality code from a single prompt, they cannot execute the code autonomously. Consequently, they do not receive feedback in the form of runtime errors or diagnostic messages. These error messages, when returned to the LLM can provide valuable information to improve the generated code. To improve their problem-solving capabilities, we embed these LLMs in agentic frameworks built with AutoGen (Python package \texttt{autogen-agentchat} v0.2.36), an open-source framework from Microsoft for building applications in which multiple AI agents converse to solve tasks~\cite{wu2023autogenenablingnextgenllm,autogen_agentchat_0_2_36}. Within this setup, we create AI agents that leverage the LLMs' capabilities to automate complex workflows, including retrieving knowledge, formulating problems, writing and executing code, analyzing results, and iteratively refining solutions.

AutoGen provides abstractions for defining agents and running one-to-one or group conversations between them, as well as a controller that decides which agent should speak next. AutoGen provides three agent classes: the Assistant Agent, the User Proxy Agent and the Group Chat Manager. The Assistant Agent is the brain in the agentic framework. It is an LLM-powered conversational agent designed to read messages, call a language model, and reply with text/code, but it does not perform actions like code execution \cite{autogen02assistant}. Each agentic framework may use a different LLM to power the Assistant Agent. In contrast, the User Proxy Agent represents the user in the system and performs tasks such as executing code and returning results or error messages \cite{autogen02userproxy}. The Group Chat Manager controls the overall flow of the interaction by broadcasting messages among agents and deciding which agent acts next \cite{autogen02groupchat}.

AutoGen also offers a model client interface that can connect to any backend that implements an OpenAI-style chat-completion API. Because AutoGen’s model clients target this API, we can attach our own models, including non-OpenAI endpoints that follow the same API format. The agentic framework is prototyped in Google Colab, and FEniCS is accessed via FEM on Colab~\cite{fem_on_colab} to provide a consistent execution environment. In the following, we describe the two different agentic frameworks used in our study: the Two-agent framework and the Multi-Agent framework.

\subsubsection{Two-agent Framework}

In the two-agent framework, a single LLM operates autonomously in a closed loop of problem-solving and self-correction driven by a structured feedback channel. We use the two-agent Coder-Executor system proposed in \cite{ni2023mechagentslargelanguagemodel}. However, in  \cite{ni2023mechagentslargelanguagemodel}, the Assistant Agent, which is responsible for writing the finite-element code, was powered by a non-fine-tuned LLM. Here, we also explore fine-tuned LLMs. The User Proxy Agent executes the code, and returns error logs that enable the LLM to perform self-correction.

\begin{figure} 
  \centering
  \includegraphics[width=\textwidth,keepaspectratio]{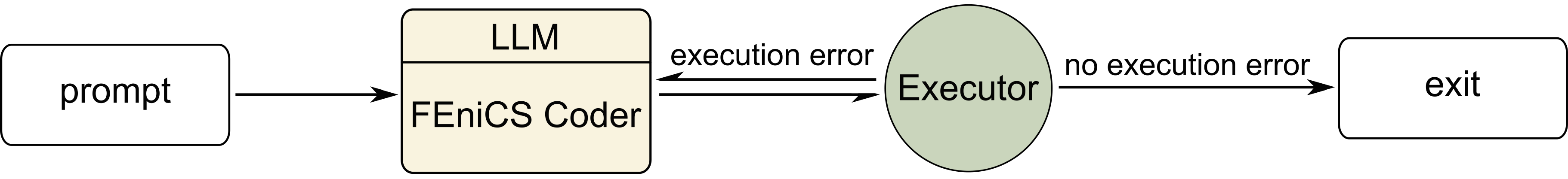}
  \caption{Two-agent framework: the FEniCS Coder (Assistant Agent) generates code from the prompt. The Executor (User Proxy Agent) runs the code, and any errors are fed back to the coder until execution succeeds.}
  \label{fig: two-agent}
\end{figure}
Fig.~\ref{fig: two-agent} shows the architecture of the two-agent framework. The framework consists of an Assistant Agent, which we call FEniCS Coder, and a User Proxy Agent, which we call Executor. The FEniCS Coder receives the full problem statement, generates FEniCS code, and forwards it to the Executor. The Executor runs the code, captures any run-time output and errors, and returns this feedback to the FEniCS Coder. The FEniCS Coder then analyzes the error messages and revises the code. This interaction continues until the code executes without errors or the LLM reaches its token limit. Because the sequence of interactions between the FEniCS Coder and Executor is fixed, a Group Chat Manager is not required. We do not intervene manually during this process. Overall, this system interprets the problems, formulates a numerical strategy, writes, debugs, and executes the FEniCS code autonomously. 

We employ two-agent frameworks in which the Assistant Agent is powered by Llama 3.2 3B, Llama 3.3 7B, GPT-OSS, Llama 3.2 3B FT, Llama 3.3 7B FT,  GPT-OSS FT and Qwen 3 32B FT.
Qwen 3 32B (non-fine-tuned) is excluded from our study because it repeatedly failed to return responses reliably through its API endpoint. 
We identify each two-agent framework by its underlying LLM; for example, the two-agent framework with Llama 3.2 3B as the Assistant Agent is denoted as Llama 3.2 3B. By deploying these LLMs in the two-agent framework, we can independently assess their capabilities and quantify the effects of model scale and fine-tuning on performance.

\subsubsection{Multi-Agent Framework}

The two-agent setup performs reasonably well overall, but it has some limitations. In the two-agent framework, a single Assistant Agent is responsible for the entire process, from formulating the problem, writing the code, and debugging it. This places a significant cognitive burden on the LLM. As the iterative feedback loop grows, the accumulation of error logs and code revisions exhausts the context window, causing the model to lose track of the critical constraints it was initially instructed to follow \cite{liu2023lostmiddlelanguagemodels}. Furthermore, this overload often causes the model to hallucinate. When the model hallucinates, the result is unpredictable. For example, the LLM may arbitrarily modify input parameters (for example, changing Young's modulus in a solid problem) \cite{zhang2025sirenssongaiocean}, leading to an incorrect solution. Another limitation of the two-agent framework is that the LLM only revises the code when it fails to compile, but a compilable code does not guarantee code correctness. Since the LLM receives no feedback on the numerical approach, it cannot correct formulation or implementation errors that still produce executable code.

\begin{figure} 
  \centering
  \includegraphics[width=\textwidth,keepaspectratio]{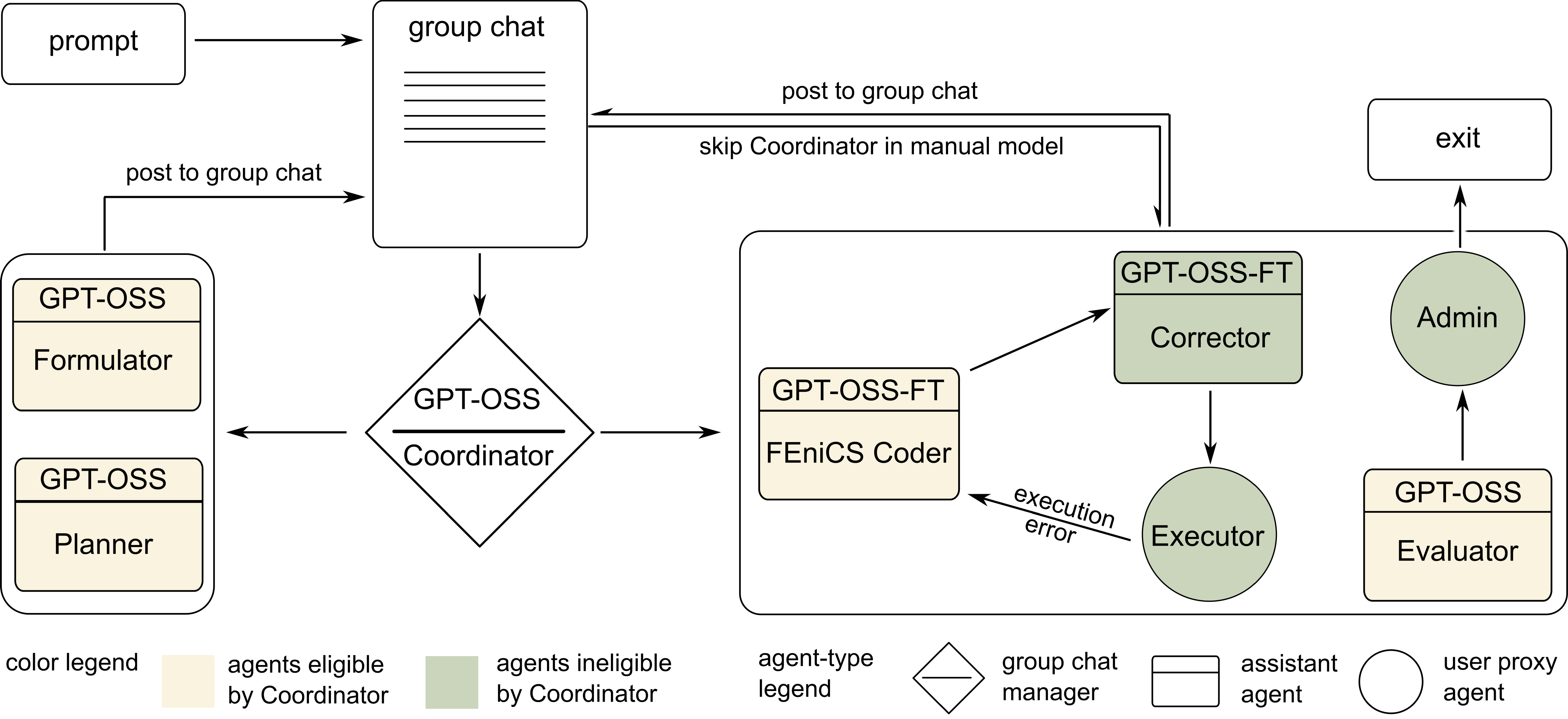}
  \caption{Schematic of the multi-agent framework. The Coordinator manages agent interactions. The agent types are identified by the type of box. For those agents powered by an LLM, we write the name of the model in their corresponding box. The yellow agents are directly eligible by the Coordinator, while the green agents are not.}
  \label{fig: multi-agent}
\end{figure}

To address these limitations, we introduce a multi-agent framework that combines five LLM-powered Assistant Agents, two User Proxy Agents, and a LLM-powered Group Chat Manager. Recent work on LLM-based multi-agent frameworks \cite{li2024agentsneed} shows that increasing the number of LLM-based agents improves performance on complex tasks through division of labor. Following this principle, our multi-agent setup assigns dedicated agents to planning, formulation, coding, execution, critique, code correction, and coordination. In the following, we describe the specific roles of each agent:

\begin{itemize} 

\item Coordinator: GPT-OSS-powered Group Chat Manager agent that is aware of the roles and responsibilities of all other agents. It oversees the group chat and intelligently selects which agent should speak next.  

\item Planner: GPT-OSS-powered Assistant Agent that proposes the overall plan for solving the problem and suggests the order of agent interactions to the Coordinator. 

\item Formulator: GPT-OSS-powered Assistant Agent that interprets the user’s prompt and formulates it into a finite-element problem specification by identifying the governing PDEs, the variational form and boundary conditions, choosing function spaces, defining the domain and mesh, and listing the outputs and post-processing required to validate the solution. 

\item FEniCS Coder: Assistant Agent powered by our fine-tuned version of GPT-OSS. Because our fine-tuned version GPT-OSS is specialized on writing FEniCS codes, this choice is expected to improve the overall performance of the agentic system. The FEniCS Coder generates legacy FEniCS codes based on the problem formulation created by the Formulator. 

\item Executor: User Proxy Agent that executes the FEniCS code and returns error messages in the group chat if the code does not run.  

\item Corrector: Assistant Agent powered by the fine-tuned version of GPT-OSS. When the FEniCS code does not execute, the Corrector examines both the FEniCS code and the error and proposes modifications for the FEniCS Coder to make it executable. 

\item Evaluator: GPT-OSS-powered Assistant Agent. The Evaluator examines the outputs of other agents in the group chat and proposes corrections to keep them aligned with the problem statement. The Evaluator provides an independent perspective on the other agents' outputs, ensuring that intermediate outputs remain coherent with the intended computational mechanics problem.

\item Admin: User Proxy Agent that enables the human user to terminate the control loop and exit the process with the current solution. 
\end{itemize}

Fig.~\ref{fig: multi-agent} shows the interactions among the agents in the multi-agent framework. First, we post the problem statement as a prompt in the group chat. The Coordinator reads the prompt and selects either the Planner, Formulator, Evaluator, or FEniCS Coder to speak next, based on the current conversation context and the information required to make progress. Although the order in which the agents participate is not fixed, the Coordinator is instructed via its system prompt to call the Planner first, typically, then the Formulator, and finally the FEniCS Coder. We have observed that, between these calls, the Coordinator often calls the Evaluator to provide a critical assessment of each agent’s output. If the Evaluator finds an output unsatisfactory, the Coordinator calls the same agent again to revise it. A distinctive feature of our framework is that an intelligent Coordinator enables a flexible, non-fixed flow of control. To illustrate this advantage, consider the following scenario: The Evaluator initially approves the problem formulation from the Formulator. Later, after the FEniCS Coder generates the code, the Evaluator realizes that the original formulation was incorrect. In such a case, a predetermined control flow would not suffice. Instead, the Coordinator recognizes the inconsistency, asks the Formulator to fix the formulation, and then resumes the workflow. 

When the Coordinator selects the FEniCS Coder, a predefined function starts a loop between the FEniCS Coder, Executor, and Corrector; see Fig.~\ref{fig: multi-agent}. The FEniCS Coder writes the code and posts it to the group chat. The Executor runs this code and returns the exit status and any error messages to the group chat. If execution fails, the Corrector inspects both the code and the error output and proposes specific modifications for the FEniCS Coder and posts them to the group chat. 
When the code executes successfully, we exit the loop and control returns to the Coordinator, which then selects the next appropriate agent. 

Although our multi-agent framework is automated, we use the Admin agent to terminate the process manually. In trials without the Admin agent, we observed that the Coordinator does not terminate the process automatically. Instead, it keeps selecting agents indefinitely. To avoid this behavior, we add the Admin agent so a human user can stop the process. 

When the Coordinator chooses the Evaluator, a predefined function hands control to the Admin after the Evaluator responds; see Fig.~\ref{fig: multi-agent}. At this point, the Admin asks the human user whether to exit the loop or continue. 
If the human user chooses to continue, control returns to the Coordinator, which then selects the next appropriate agent. 
If the human user exits the loop, the process terminates, and the current code and solution generated by the framework are stored for evaluation.

Importantly, the human user does not decide on termination based on the quality of the solution. Instead, the human user exits the loop only when the Evaluator is satisfied with the solution. 
Otherwise, the human user continues the loop and hands back control to the Coordinator. Thus, the multi-agent framework remains automated, and the manual intervention serves only to compensate for the Coordinator's inability to exit the loop.

We assign roles to the agents by specifying their role descriptions in the AutoGen framework. In long conversations, agents may hallucinate and forget their assigned roles \cite{zhang2025sirenssongaiocean, chang2023surveyevaluationlargelanguage, agarwal2025codemiragehallucinationscodegenerated}. To reduce the risk of hallucination, we reinforce role assignment using a separate LLM for each assistant agent. Each LLM instance carries a dedicated system prompt that specifies its own responsibilities and summarizes the roles of the other agents it is interacting with.

\section{Evaluation Methodology and Results}

\subsection{Evaluation Methodology}

In this section, we compare the performance of the different agentic frameworks. To benchmark our agentic frameworks, we will also compare them against GPT-5 Thinking. 
To emulate the behavior of computational mechanicians that are untrained on agentic AI, we deploy GPT-5 using a non-agentic setup with two zero-shot attempts. In this framework, the Thinking version of GPT-5 is first prompted with the question and asked to produce a FEniCS code to solve it. If GPT-5 produces non-executable code on the first attempt, it is prompted again in a fresh session with no memory retention. The code from this second prompt is taken as final and used for evaluation.

\paragraph{Test set}

We evaluate the capability of our agentic frameworks by testing them on a set of 39 computational mechanics problems. This benchmark set was designed by domain experts in computational mechanics. All 39 problems are unique and separate from the datasets used for training the LLMs. To assess the distinction between the evaluation problems and the training data, we compared the evaluation problems against the training set using detectable attributes (PDE family, geometry of the domain, and boundary-condition structure, including weak-form boundary-integral markers). The 1005 problems in the training dataset are heavily skewed toward core PDE classes on canonical 2D geometries (e.g., Poisson or heat equations on unit-square–type domains) with frequent Dirichlet and common weak-form Neumann or Robin terms, and relatively few periodic boundaries or interface conditions. In contrast, the test set focuses on coupled and higher-complexity systems, includes no standalone Poisson or heat equations, and contains 5/39 questions with PDE families or couplings absent from training. Some examples of PDE families or couplings entirely missing in the training set include the Allen-Cahn~\cite{allen1979microscopic} equation, a Boussinesq-type Navier-Stokes-temperature coupling~\cite{chandrasekhar1961hydrodynamic}, Brinkman flow/free-flow coupling, and Stokes flow/Darcy flow coupling \cite{beavers1967boundary, saffman1971boundary}. Even where component PDEs exist in training, several evaluation items are novel combinations (e.g., Navier-Stokes coupled with advection-diffusion, or flow coupled with a nonlinear solid) that are essentially absent from the training distribution. The evaluation benchmarks include problems of linear and nonlinear elasticity, plasticity, Newtonian and non-Newtonian flow, fluid structure interaction, reaction-diffusion problems on moving domains, and other multiphysics applications. For evaluation purposes, the set of problems was split into three subsets using two different criteria. The first criterion refers to the physics solved and consists of solid mechanics problems (16), fluid mechanics problems (15), and multiphysics problems (8). The second criterion is difficulty and includes easy (13), medium (13), and hard (13) problems, where the difficulty levels were assigned by domain experts. Each problem specifies the geometry, the mathematical model, the boundary conditions, and the expected output. The expected output is the primary unknown associated with the problem. For example, the expected solutions for solid mechanics and fluid mechanics problems are the displacement and velocity, respectively. The full list of problems is provided in the Appendix ~\ref{subsec:benchmark}. Here, we present three representative questions from the test set: an easy solid mechanics problem (\hyperref[box:prob1]{Sample Problem 1}), a medium fluid mechanics problem (\hyperref[box:prob2]{Sample Problem 2}) and a hard multiphysics problem (\hyperref[box:prob3]{Sample Problem 3}). For better readability, the problems are shown here in LaTeX fonts, but they were prompted to the agentic systems as plain text with Unicode symbols.

\begin{tcolorbox}[enhanced,
breakable,colback=gray!10,colframe=gray!50,title={Sample Problem 1: Easy, Solid Mechanics},label=easy-solid,  
    phantom={\phantomsection\label{box:prob1}}]
\textbf{Geometry:} \\
Let $\Omega = (0, 1.0) \times (0, 0.20)$~m be a rectangular plate. \\

\textbf{Model:} \\
Plane-stress linear elasticity for displacement $u = (u_x, u_y)$ in $\Omega$. \\
$\sigma$ is the Cauchy stress tensor and $n$ is the outward unit normal. \\

\textbf{Material:} \\
Young's modulus $E = 200$~GPa, Poisson's ratio $\nu = 0.30$. \\

\textbf{Boundary conditions:} \\
Left edge ($x = 0$): fixed, $u_x = 0$, $u_y = 0$. \\
Right edge ($x = 1$): prescribed displacement, $u_x = 0.001$~m, $u_y = 0$. \\
Top ($y = 0.20$) and bottom ($y = 0$): traction-free ($\sigma n = 0$). \\

\textbf{Mesh:}  \\
Uniform structured mesh with $20 \times 4$ subdivisions across $(x, y)$. \\

\textbf{Output:} \\
Save a color map of the horizontal displacement $u_x$ as \texttt{q1\_ux.png}. \\
Save the resulting displacement field in XDMF format. 
\end{tcolorbox}

\begin{tcolorbox}[enhanced,
breakable,colback=gray!10,colframe=gray!50,title={Sample Problem 2: Medium, Fluid Mechanics} , label=med-fluid, 
    phantom={\phantomsection\label{box:prob2}}]
\noindent \textbf{Geometry:} \\
Let $\Omega = [0, 1] \times [0, 1]$~m be a unit square cavity.

\medskip
\noindent \textbf{Model (Boussinesq, steady):} \\
Solve the steady incompressible Navier-Stokes with body force $f$ coupled to a steady advection-diffusion equation for temperature $T$. \\
The notation for velocity is $u = (u_x, u_y)$, pressure is $p$, and temperature is $T$. \\
The body force $f$ is modeled as a Boussinesq buoyancy term, $f = [0, \rho g \beta (T - T_{\text{ref}})]$, where $\rho$ is fluid density, $g$ is the gravitational acceleration, $\beta$ is the volumetric thermal expansion coefficient, $T_{\text{ref}}$ is a reference temperature.

\medskip
\noindent \textbf{Boundary conditions:} \\
Left wall ($x = 0$): $T = 1$ (hot), no-slip and no penetration. \\
Right wall ($x = 1$): $T = 0$ (cold), no-slip and no penetration. \\
Top/bottom ($y = 1$ and $y = 0$): adiabatic $\partial T/\partial n = 0$, no-slip and no penetration. \\
$n$ is the outward unit normal vector.

\medskip
\noindent \textbf{Parameters:} \\
Density $\rho = 1$~kg/m$^3$. \\
Dynamic viscosity ($\mu$): $1.5 \times 10^{-5}$~Pa$\cdot$s. \\
Thermal diffusivity $\alpha = 2.1 \times 10^{-5}$~m$^2$~s$^{-1}$. \\
$g \beta = 3.15 \times 10^{-5}$~m~s$^{-2}$~K$^{-1}$. \\
$T_{\text{ref}} = 0.5$~K.

\medskip
\noindent \textbf{Output:} \\
Temperature field: save a color map as \texttt{q11\_T.png}. \\
Report the average Nusselt number at the left wall. \\
Also, save the velocity field ($u$), pressure field ($p$), and temperature field ($T$) to \texttt{q11\_solution.xdmf}.

\end{tcolorbox}

\begin{tcolorbox}[enhanced,
breakable,colback=gray!10,colframe=gray!50,title={Sample Problem 3: Hard, Multiphysics}, label=hard-multi, 
    phantom={\phantomsection\label{box:prob3}}]
Analyze the motion of a solid flag in a 2D flow channel using a fluid structure interaction model. \\

\noindent \textbf{Geometry:} \\
The fluid domain is a 2D channel of length $2.5$~m and height $0.41$~m. The flag is modeled as a rectangular elastic solid of length $0.35$~m and thickness $0.02$~m with the right-bottom corner placed at $(0.60 \text{ m}, 0.19 \text{ m})$ in the reference configuration. \\[.2cm]
The pole of the flag is modeled as circular disk of radius $0.05$~m centered at $(0.20 \text{ m}, 0.20 \text{ m})$. The pole is assumed to be rigid and we remove this disk from our computational domain. Since this disk intersects the fluid and solid domain, portions of both the solid and fluid domain are removed.  \\[.2cm]
The boundary of the circular disk is denoted by $\Gamma = \Gamma_s \cup \Gamma_f$, where $\Gamma_f$ and $\Gamma_s$ are the partitions of $\Gamma$ that are in contact with the fluid and solid domain, respectively.

\medskip
\noindent \textbf{Model:} \\
The fluid is incompressible and is solved in the moving domain using the Arbitrary Lagrangian Eulerian form of the Navier-Stokes equations. The solid flag is modeled using the St. Venant-Kirchhoff model. \\[.2cm]
Enforce no-slip (fluid velocity equals solid velocity) and traction balance (fluid traction equals solid traction) at the fluid-solid interface.

\medskip
\noindent \textbf{Boundary conditions:} \\
At the inlet ($x = 0$), prescribe a parabolic velocity profile with: \\
$u(t, 0, y) = \begin{cases} u_y(0,y)  \frac{1 - \cos(\pi t / 2)}{2}, & \text{if } t < 2.0 \text{ sec.} \\ u_y(0,y), & \text{if } t \ge 2.0 \text{ sec.} \end{cases}$ \\
where $u_y(0,y) = 1.5 \bar{U}  y (H - y) / (H/2)^2$ with $H = 0.41$ and $\bar{U} = 1$~m/s. \\
At the outlet ($x = 2.5$~m), use a traction free boundary condition. \\
Impose no-slip and no penetration on the top and bottom channel walls. \\
At $\Gamma_f$: Impose no slip and no penetration for the fluid. \\
At $\Gamma_s$: Impose zero displacement for the solid flag.

\medskip
\noindent \textbf{Parameters:} \\
Use fluid density $\rho_f = 1000$~kg/m$^3$ and kinematic viscosity $\nu_f = 1.0\times10^{-3}$~m$^2$/s. Use solid density $\rho_s = 10000$~kg/m$^3$, Poisson ratio $\nu_s = 0.4$, shear modulus $\mu_s = 0.5\times10^6$~Pa.

\medskip
\noindent \textbf{Numerical outputs:} \\
Save the fluid velocity and pressure fields and the beam displacement in XDMF format. \\
Report the displacement components of point A with time, where the reference configuration of point A is given by $A(t=0) = (0.60 \text{ m}, 0.20 \text{ m})$.
\end{tcolorbox}

\begin{figure} 
  \centering
  \includegraphics[width=\textwidth,keepaspectratio]{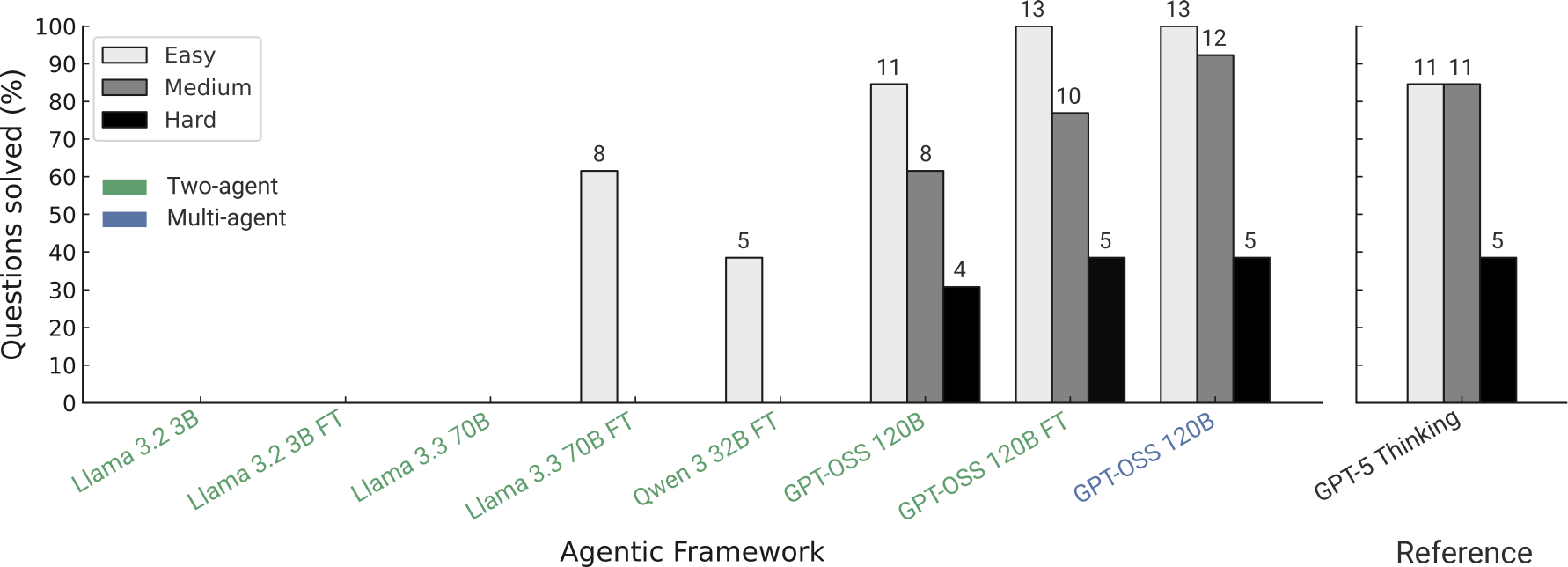}
  \caption{Performance comparison of agentic LLM frameworks and GPT-5 across difficulty levels. Bars represent the percentage of questions solved correctly, while numerals over bar indicate the number of questions solved correctly.}
  \label{fig: Results_by_difficulty}
\end{figure}

\paragraph{Code correctness evaluation} \label{correct}

To evaluate the performance of the agentic frameworks, we prepare a reference solution for each question using our domain expertise. If the solution produced by the framework matches the reference solution, we mark it as correct. To perform the evaluation, we followed a multi-step solution comparison process against reference solutions. First, we labelled as incorrect all non-executable codes, while executable codes were passed on to the next evaluation steps. Second, we plot the reference solution and the agentic framework solution on the same plot and consider that they match if they are indistinguishable at the scale of the plot. We also compare both solutions along a characteristic line in the domain using standardized axes and use the same visual criterion. Finally, we utilized an LLM-judge to identify discrepancies between the reference code and the agentic framework's solution code followed by manually verifying the differences marked by LLM-judge. This step ensures that the underlying physical formulations are consistent and prevents the misclassification of incorrect implementations that may appear identical to the reference at the plotted scale in the second step. For example, in first Solid Mechanics problem 1, Llama 3.3 70B (two-agent framework) generated a solution plot indistinguishable from the reference (see Fig.~\ref{fig: solid1}). However, a code-level analysis using an LLM-judge revealed that the framework employed a plane strain formulation instead of the plane stress formulation required by the reference solution.  Therefore, we marked the solution as incorrect.

We considered alternative correctness criteria, including a maximum relative $L^2$ error with respect to the ground truth solution. However, we did not use a relative $L^2$ error as the primary metric because different agentic frameworks can produce solutions on different meshes. Computing an $L^2$ error would require transferring one solution to a common finite-element space, which is difficult to automate robustly at scale and introduces projection errors that can obscure true discrepancies. To show that our multi-step solution comparison evaluation methodology is robust, we performed a focused comparison between our criterion and $L^2$-error-based evaluation on five representative problems, including threshold sensitivity; see Appendix~\ref{app}. 

\paragraph{Performance metric}

We quantify the performance of each agentic framework using accuracy, defined as the number of correctly solved questions divided by the total number of questions. Fig.~\ref{fig: Results_by_difficulty} shows the performance of our agentic frameworks across the different levels of difficulty, depicting the results of the two-agent frameworks with green fonts and the results of the multi-agent framework with blue fonts. The agentic frameworks are identified by the name of the LLM that writes the code. For reference, we also include the results of GPT-5 Thinking (black fonts). Fig.~\ref{fig: Results_by_subject} shows the results of the two-agent frameworks, the multi-agent framework, and GPT-5 grouped by physics category.

\subsection{Results}\label{subsec:results}
In this section, we report the performance of different two-agent frameworks, Multi-agent framework and GPT-5 Thinking on the Test set containing 39 benchmark problems. Plots of the numerical solutions generated by agentic frameworks and GPT-5 Thinking to some of the problems in the Test set are presented in Appendix ~\ref{appc}. 

In what follows, we first compare the performance of the two-agent frameworks. This comparison allows to assess the capabilities of different LLMs independently and quantify the effects of model scale and fine-tuning on performance.

\paragraph{Model Size Improves Performance in a Two-agent Framework} 
The two-agent frameworks with the smallest models in the study, i.e., Llama 3.2 3B, and its fine-tuned version, fail to produce executable FEniCS code for any of the problems tested. Llama 3.3 70B produces executable code for 9 problems in the solid mechanics set, but does not solve any of them correctly. GPT-OSS 120B performs substantially better, achieving an overall accuracy across all 39 test problems of 58.97\%. From these results of the non-fine-tuned two-agent frameworks, we observe that increasing model size leads to better performance. 
A similar trend is apparent when we compare the fine-tuned two-agent frameworks. The accuracy across all 39 test problems of GPT-OSS 120 B FT (71.79 \%) exceeds that of Llama 3.3 70B FT (20.51 \%), which is greater than Qwen 3 32B FT (12.82 \%). 

The effect of model size is especially pronounced as problem difficulty increases.   
Llama 3.3 70B FT and Qwen 3 32B FT fail to solve any problems on the medium and hard question set; see Fig~\ref{fig: Results_by_difficulty}. In contrast, GPT-OSS 120B and GPT-OSS 120B FT perform significantly better than the smaller models on the medium- and hard-question sets. The mixture-of-experts and thinking design of GPT-OSS help it handle the multi-step reasoning and debugging required to solve these problems more effectively than the non-thinking Llama 3.3 70B.

\begin{figure} 
  \centering
  \includegraphics[width=\textwidth,keepaspectratio]{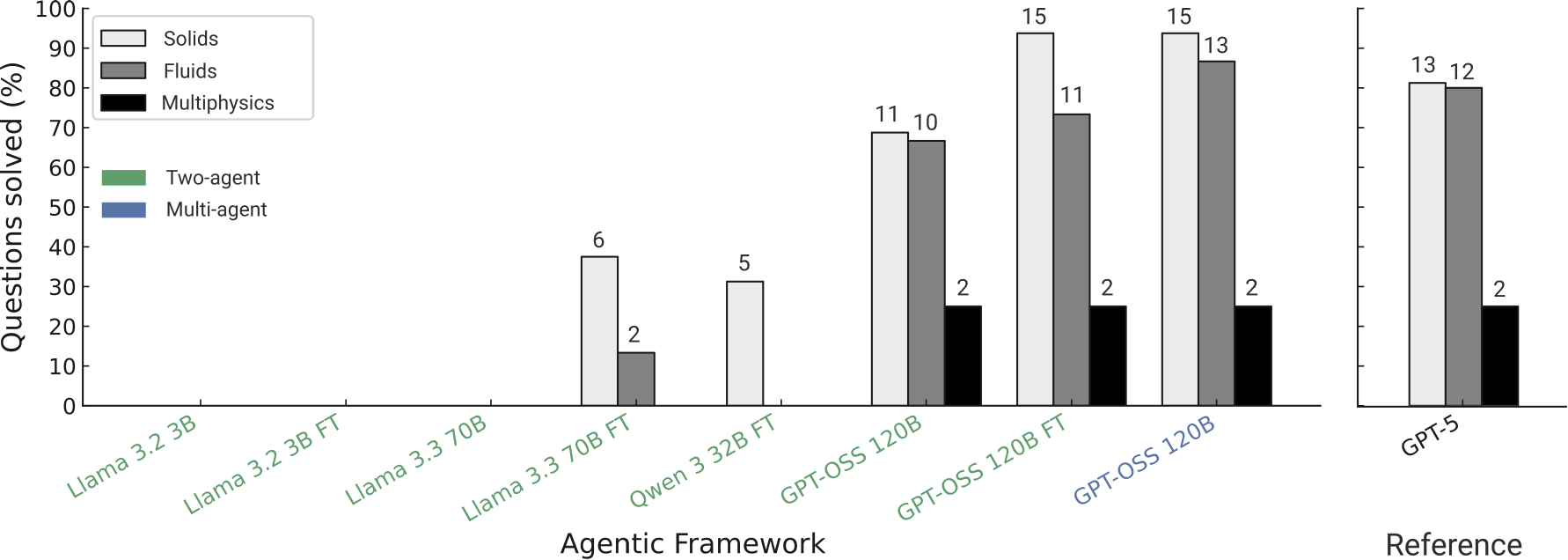}
  \caption{Performance comparison of agentic LLM frameworks and GPT-5 across solid, fluid, and multiphysics questions. Bars represent the percentage of questions solved correctly, while numerals over bar indicate the number of questions solved correctly.}
  \label{fig: Results_by_subject}
\end{figure}

Fig.~\ref{fig: Results_by_subject} shows that increasing model size also improves performance across all physics categories. For example, GPT-OSS 120B FT solves 15 solid, 11 fluid, and 2 multiphysics questions, whereas Llama 3.3 70B solves only 6 solid, 2 fluid, and no multiphysics questions.  
\paragraph{Fine-tuning Improves Performance in a Two-agent Framework} 
From Fig.~\ref{fig: Results_by_difficulty}, we observe that fine-tuning LLMs on our FEniCS dataset yields consistent gains across all two-agent frameworks.  
Small models benefit the most from fine-tuning. The gains are most evident for Llama 3.3 70B, whose fine-tuned version correctly solves 8 problems, whereas the non-fine-tuned version solves none. Fine-tuning also enables Qwen 3 32B FT to outperform the larger Llama 3.3 70B. 

GPT-OSS, our largest fine-tuned model, also benefits from fine-tuning: it solves more solid and fluid problems (Fig.~\ref{fig: Results_by_subject}) and outperforms the non-fine-tuned model across the easy, medium, and hard sets (Fig.~\ref{fig: Results_by_difficulty}). 

To further understand the impact of fine-tuning, we compare the answers of Llama 3.3 70B and its fine-tuned version on \hyperref[box:prob1]{Sample Problem 1}. This problem requires the solver to compute the expansion of a 2D rectangular solid plate using a plane-stress linear elasticity model. \hyperref[box:block1]{Code block 1} shows the solution produced by the base Llama 3.3 70B model, whereas \hyperref[box:block2]{Code block 2 } shows the solution produced by the fine-tuned model. Llama 3.3 70B incorrectly solves this problem by assuming plane strain, whereas the fine-tuned version correctly uses plane stress and produces the correct solution. In this case, fine-tuning improves the model by helping it correctly interpret and apply the plane-stress assumption.

\begin{tcolorbox}[
enhanced,
breakable,
    colframe=gray!60,
    colbacktitle=gray!60,
    coltitle=white,
    colback=LightGray,
    fonttitle=\small,
    arc=1pt, boxrule=1pt,
    title={\centering Block 1 - Llama 3.3 3B non-fine-tuned - Sample Problem 1}, label=block1, 
    phantom={\phantomsection\label{box:block1}},
    left=1pt, right=1pt, top=0pt, bottom=0pt, boxsep=1pt
]
\begin{lstlisting}[xleftmargin=2mm]
# Define stress and strain tensors
def sigma(u, E, nu):
  mu = E / (2 * (1 + nu))
  lmbda = E * nu / ((1 + nu) * (1 - 2 * nu))
  return 2 * mu * sym(grad(u)) + lmbda * tr(sym(grad(u))) * Identity(2)
def epsilon(v):
  return sym(grad(v))
a = inner(sigma(u, E, nu), epsilon(v)) * dx
l = inner(Constant((0, 0)), v) * ds
\end{lstlisting}
\end{tcolorbox}

\begin{tcolorbox}[
enhanced,
breakable,
    colframe=gray!60,
    colbacktitle=gray!60,
    coltitle=white,
    colback=LightGray,
    fonttitle=\small,
    arc=1pt, boxrule=1pt,
    title={\centering Block 2 - Llama 3.3 3B fine-tuned - Sample Problem 1},label=block2, 
    phantom={\phantomsection\label{box:block2}}, 
    left=1pt, right=1pt, top=0pt, bottom=0pt, boxsep=1pt
]
\begin{lstlisting}[xleftmargin=2mm]
# Plane stress elasticity tensor
mu = E/(2.0*(1.0+nu))
lmbda = E*nu/((1.0+nu)*(1.0-2.0*nu)) # not used directly for plane stress

# Plane stress constitutive matrix (Voigt notation)
def sigma(eps):
  # eps is a 2x2 symmetric tensor
  eps_xx = eps[0,0]
  eps_yy = eps[1,1]
  eps_xy = eps[0,1]
  factor = E/(1.0 - nu**2)
  s_xx = factor*(eps_xx + nu*eps_yy)
  s_yy = factor*(nu*eps_xx + eps_yy)
  s_xy = factor*((1.0 - nu)/2.0)*2.0*eps_xy
  return as_tensor([[s_xx, s_xy],
  [s_xy, s_yy]])
\end{lstlisting}
\end{tcolorbox}

\paragraph{GPT-OSS FT in a Two-agent Framework Outperforms a non-agentic GPT-5 Thinking Deployment}
Across all levels of difficulty, GPT-OSS FT performs better than GPT-5 Thinking, despite using far fewer parameters and lower inference cost. This comparison demonstrates that domain-specific fine-tuning, combined with a two-agent execution-feedback loop, can be a viable open-source, lower-cost alternative to a much larger proprietary model on FEniCS-based computational mechanics problems. 

\paragraph{Multi-agent Framework Performs Better than Two-agent Framework}
To understand the improvement provided by the multi-agent framework, we compare the performance of the multi-agent framework with the two-agent framework GPT-OSS-FT. As shown in Fig.~\ref{fig: Results_by_difficulty}, the multi-agent framework has solved two more questions than the two-agent GPT-OSS-FT in the medium difficulty set, while both have the same accuracy in the easy and hard sets. Although they have the same accuracy in the hard problem set, the multi-agent framework produces better code than the two-agent GPT-OSS-FT.
To illustrate this, we compare the answers for \hyperref[box:prob4]{Sample Problem 4}. This problem asks the models to solve the transport of a chemical species inside an expanding circular disk using an Arbitrary Lagrangian–Eulerian (ALE) formulation. The multi-agent framework correctly implements the ALE form of the chemical transport equation, but fails to impose the boundary conditions accurately and therefore does not answer the question correctly. The two-agent GPT-OSS-FT fails to correctly implement the ALE formulation as well as the boundary conditions. \hyperref[box:block3]{Code block 3} shows the implementation produced by the two-agent GPT-OSS-FT, whereas \hyperref[box:block4]{Code block 4} shows the implementation produced by the multi-agent framework; together, they illustrate that the multi-agent framework captures the ALE formulation more accurately, even though both models answer the question incorrectly. 
\begin{tcolorbox}[enhanced,breakable,colback=gray!10,colframe=gray!50,title={Sample Problem 4}, label=hard-multi, 
    phantom={\phantomsection\label{box:prob4}}]
\textbf{Geometry: }\\
Solve the transport of a chemical inside an expanding circular disk $\Omega(t)$ with radius $R(t)$, where $R(t) = R_0 + s \cdot t$ with constant rate $s = 1.0 \times 10^{-4}$ m s$^{-1}$. Here, $R_0$ is the initial radius and is equal to 0.05 m. The boundary of this circular disk is denoted by $\Gamma(t)$. \\

\textbf{Model: }\\
Solve the diffusion-reaction equation for concentration $c$ on a moving domain using an Arbitrary Lagrangian-Eulerian description, with constant diffusivity $D$ and first-order decay equal to $\kappa c$, where $\kappa$ is the decay rate and is positive. \\

\textbf{Mesh:} \\
Unstructured triangular mesh of $\Omega(0)$ with characteristic size $h_0 \approx 1.0 \times 10^{-3}$ m. \\ 
The mesh motion is given by $w(x,t) = s x / \|x\|$ for 
$x \neq (0,0)$, $w(x,t) = 0$ for $x = 0$. \\

\textbf{Boundary conditions:} \\
Zero total flux at the moving boundary: $(-D \nabla c - w c) \cdot n = 0$ on $\Gamma(t)$ for all $t \ge 0$, where $D$ is the diffusion coefficient. \\

\textbf{Initial condition:}\\
$c(x,0) = 1$ for $x \in \Omega(0)$. \\

\textbf{Parameters:} \\
Diffusivity: $D = 1.0 \times 10^{-5}$ m$^2$ s$^{-1}$. \\
Decay rate: $\kappa = 1.0 \times 10^{-4}$ s$^{-1}$. \\
Use the time step $\Delta t = 0.01$ s. Simulate up to 10 seconds \\

\textbf{Output:} \\
Save the output concentration as function of time in an xdmf file. \\
Report the total concentration in the domain after every 100 time steps. \\
\end{tcolorbox}

\begin{tcolorbox}[
    colframe=gray!60,
    colbacktitle=gray!60,
    coltitle=white,
    colback=LightGray,
    fonttitle=\small,
    arc=1pt, boxrule=1pt,
    title={\centering Block 3 - GPT-OSS FT in a two-agent framework - Sample Problem 4},label=block2, 
    phantom={\phantomsection\label{box:block3}}, 
    left=1pt, right=1pt, top=0pt, bottom=0pt, boxsep=1pt
]
\begin{lstlisting}[xleftmargin=2mm]
w_vec = w.vector()
disp[mask,0] = s*dt * x[mask,0] / r[mask]
disp[mask,1] = s*dt * x[mask,1] / r[mask]
w_vec.set_local(disp.ravel())
# ---- move mesh ----
ALE.move(mesh, w)
# ---- update function space after mesh move ----
V = FunctionSpace(mesh, 'Lagrange', 1)
c_n = interpolate(c_n, V)  # project previous solution onto new mesh
# ---- variational problem (backward Euler, ALE) ----
c = TrialFunction(V)
v = TestFunction(V)
a = (c*v + dt*D*dot(grad(c), grad(v)) + dt*kappa*c*v)*dx
L = (c_n*v - dt*dot(w, grad(c_n))*v)*dx
\end{lstlisting}
\end{tcolorbox}

\begin{tcolorbox}[
    colframe=gray!60,
    colbacktitle=gray!60,
    coltitle=white,
    colback=LightGray,
    fonttitle=\small,
    arc=1pt, boxrule=1pt,
    title={\centering Block 4 - GPT-OSS FT in a multi-agent framework - Sample Problem 4},label=block2, 
    phantom={\phantomsection\label{box:block4}}, 
    left=1pt, right=1pt, top=0pt, bottom=0pt, boxsep=1pt
]
\begin{lstlisting}[xleftmargin=2mm]
# --- mesh velocity on current mesh ---
w = interpolate(w_expr, Vv)
# --- displacement over one time step ---
disp = Function(Vv)
disp.vector()[:] = w.vector() * dt
# --- move mesh ---
ALE.move(mesh, disp)
# --- interpolate previous solution onto new mesh ---
c_n = interpolate(c_n, V)

# --- trial / test functions on updated space ---
c = TrialFunction(V)
v = TestFunction(V)
# --- variational forms (Backward Euler) ---
a = (c*v/dt + D*dot(grad(c), grad(v)) + kappa*c*v + dot(w, grad(v))*c)*dx
L = (c_n*v/dt)*dx
# --- solve for new concentration ---
c = Function(V)
solve(a == L, c)
\end{lstlisting}
\end{tcolorbox}

\paragraph{Limitations of the framework}

Building the ALL-FEM dataset, agentic process, and assessment methodology requires various design choices that restrict the scope of the study and the interpretation of the results. Some of these limitations are listed below.

First, our benchmark prompts do not specify mesh details (e.g., target resolution, element type, or discretization approach), which implies that different agentic frameworks may solve the same problem on different meshes. While this renders agentic systems that are potentially more useful, it makes it difficult to apply a single quantitative error metric (e.g., the relative $L^2$ error) consistently across the full benchmark. The reason is that computing $L^2$ errors between solutions on non-matching finite-element spaces requires a projection, which in general introduces error and complicates the choice of a uniform pass/fail threshold. As a result, we decided to use a standardized multi-step solution comparison process performed by domain experts; see Section ~\ref{correct}. Although this remains a limitation of our evaluation protocol, a detailed quantitative comparison with an $L^2$-based evaluation shows that it is a robust and reliable approach; see Appendix~\ref{app}.

Second, the multi-agent system is not totally autonomous end-to-end since it presently depends on a human termination step via the Admin agent to avoid uncontrolled execution cycles. In our studies, the human did not give problem-solving help or corrections; the sole action was to stop the run after the Evaluator showed that the answer passed the acceptance requirement.

Lastly, the benchmark design concentrates on solver-generation and verification on basic canonical geometries and does not fully capture the complexities of real-world engineering processes. 
Our models are fine-tuned for code generation, rather than for robust automated meshing or tagging. By utilizing relatively simple 2D geometries, we ensure that any failures are clearly attributed to the solver-side formulation rather than to meshing or tagging errors. However, this simplification can under-represent challenges such as 3D meshing or performance under incomplete tags. For instance, in Multiphysics Problem 6 (the Turek–Hron FSI benchmark~\cite{turek2006proposal}), the agents did not implement adequate mesh refinement around small geometric features, resulting in an unsatisfactory representation of some boundaries. While standard workflows often rely on external CAD-to-mesh pipelines~\cite{geuzaine2009gmsh}, a potential approach for end-to-end autonomy would be to couple our framework with a dedicated geometry-and-meshing pipeline, such as STEP-LLM~\cite{shi2026stepllm}. To support a clearer interpretation of these numerical results, we report the mesh statistics used for each benchmark in Appendix~\ref{app-mesh} (Tables~\ref{tab:mesh_solid}--\ref{tab:mesh_multiphysics}).


\section{Conclusions}
We have developed ALL-FEM, an autonomous simulation system that combines agentic AI and fine-tuned LLMs to automate substantial portions of the finite-element workflow using FEniCS. Starting from a curated seed dataset of over 500 expert-generated FEniCS codes and using a novel pipeline for synthetic code generation, we developed a dataset with over 1000 entries. We leveraged this dataset to fine-tune LLMs spanning 3B-120B parameters using LoRA/QLoRA. We embedded off-the-shelf LLMs and our fine-tuned models in a minimal two-agent Coder-Executor loop and in a richer multi-agent system. 

We evaluated our agentic systems using a benchmark composed of 39 problems that include linear/nonlinear elasticity, plasticity, Newtonian/non-Newtonian flow, thermofluids, fluid-structure interaction, phase separation, and transport on moving domains. We found that the flow automation achieved by our multi-agent framework allowed the LLMs to receive execution feedback in the form of runtime errors, which improved their performance. Our results also showed that fine-tuning consistently improves code correctness and demonstrates that domain-specific fine-tuning can transform general-purpose LLMs into specialized solvers capable of accurately formulating computational mechanics problems and implementing them in FEniCS. Notably, we achieved a very high performance gain from fine-tuning even though our training data was relatively small. This trend suggests that further scaling the dataset could yield additional improvements in robustness and accuracy of the codes generated. 

Our fine-tuned version of GPT-OSS 120B embedded in a multi-agent framework achieved the highest performance. Within the limits of our setup, this configuration also surpasses a non-agentic deployment of GPT-5 Thinking, showing that fine-tuned, small models used in a suitable agentic system can rival or even outperform much larger proprietary systems used in naive ways.


This work makes clear that a reliable and automatic simulation system is not yet possible, and identifies gaps where further research is warranted. Future work should prioritize collaboration with the global finite element community to aggregate a comprehensive code database. Developing such a large-scale resource would allow for the training of highly specialized models, thereby creating a more robust framework for fully automating the engineering workflow. Our own contribution in this direction focuses on creating {\tt FEniCS-LLM}, an open source LLM based on open data that outperforms existing LLMs in FEniCS code generation; see \href{https://fenics-llm.github.io}{https://fenics-llm.github.io}. Another critical step toward automated simulation systems is to incorporate a wider range of agents, including result analyzers that leverage the visual recognition capabilities of LLMs to analyze results, which would enable the framework to produce more reliable codes and give it a better ability to correct implementation errors.

Overall, this research shows that combining curated domain data, parameter-efficient fine-tuning, and agentic orchestration is a feasible approach for developing autonomous simulation systems in computational mechanics. Our findings further suggest that this approach may extend to other scientific and engineering domains with similarly intricate, software-dependent workflows.

\section{Acknowledgements} 
This work was partially supported by Eli Lilly and Company  (United States).
The opinions, findings, and conclusions or recommendations expressed are those of the authors and do not necessarily reflect the views of Eli Lilly and Company.
This work uses the Bridges-2 system at the Pittsburgh Supercomputing Center (PSC) through allocation no. MCH220014 from the Advanced Cyberinfrastructure Coordination Ecosystem: Services and Support (ACCESS) program, which is supported by the National Science Foundation, grant nos. 2138259, 2138286, 2138307, 2137603, and 2138296.

\begin{appendices}
\section{Code correctness criterion based on the $L^2$ norm} \label{app}


To evaluate ALL-FEM, we checked code correctness using a multi-step solution comparison process; see Section ~\ref{correct}.
Here, we test the robustness and accuracy of our process by comparing it with quantitative $L^2$-error metrics. We perform two $L^2$-error-based validation tests on five representative problems and compare their pass/fail outcomes against our multistep comparison process in Tables~\ref{tab:thresholds_updated}--\ref{tab:thresholds_test2_updated}.

In Test 1, we compute a reference solution on a mesh that is finer than the meshes used by all framework-generated solutions. We then interpolate each framework solution onto this reference mesh~\cite{fenics_nonmatching_interpolation_demo_2017} so that the reference and generated fields are defined on the same discretization. On this common mesh, we compute the relative $L^2$ error for each primary field. A framework output is labeled ``correct'' if its relative $L^2$ error is below a chosen threshold, and we sweep this threshold from 1\% to 4\% to show how stricter accuracy requirements affect pass rates. Table~\ref{tab:thresholds_updated} reports the resulting pass counts and average errors across thresholds and shows that, for these representative cases, a 4\% threshold yields pass or fail decisions that are almost consistent with multi-step solution comparison evaluation process. 

In Test 2, we remove the interpolation step to determine whether the threshold sensitivity in Test 1 is mainly an artifact of transferring solutions between meshes. For each agentic output, we construct a matched reference solution that uses the same mesh, same finite-element spaces, and the same discretization settings as the framework-generated code, while keeping the governing equations, boundary and initial conditions, and physical parameters identical to the original reference formulation. This places the reference and generated solutions in the same finite-element space, so the $L^2$ error is computed directly without projection or interpolation mismatch. We then label an output as correct if its relative $L^2$ error is below 0.001\% and compare these labels to the multi-step solution comparison evaluation process in Table~\ref{tab:thresholds_test2_updated}. The two evaluation methods yield identical labels across all cases. The average errors among passing cases are substantially smaller than in Test 1, indicating that mesh-to-mesh transfer can inflate apparent discrepancies. Therefore, Test~2 provides the cleanest quantitative assessment, but it requires extensive manual intervention for each agentic output and does not scale well. 
Considering this scalability constraint and the agreement between our multi-step solution comparison process for labelling and the same-mesh $L^2$ labelling in Test~2, we decided to use the multi-step solution comparison process for evaluation.\\

\begin{table}[h]
\centering
\small
\caption{Test 1: Sensitivity of pass counts and average $L^2$ error norms compared to multi-step solution comparison evaluation process.}
\label{tab:thresholds_updated}
\setlength{\tabcolsep}{3pt} 
\renewcommand{\arraystretch}{1.5} 

\begin{tabularx}{\linewidth}{l ccc >{\centering\arraybackslash}X >{\centering\arraybackslash}p{2.2cm}}
\toprule
\textbf{Validation case} & \multicolumn{3}{c}{\textbf{\makecell{Pass count\\($L^2$ Threshold)}}} & \textbf{\makecell{Avg. Relative\\$L^2$ norm (\%)\\(Passed, 4\%)}} & \textbf{\makecell{Multi-step\\solution\\comparison\\evaluation\\process}} \\
\cmidrule(lr){2-4}
 & \textbf{4\%} & \textbf{2\%} & \textbf{1\%} & & \\
\midrule
\multicolumn{6}{l}{\textit{\textbf{Solid Mechanics}}} \\
Solid Q4  & 6/7 & 4/7 & 4/7 & $4.41 \times 10^{-1}$ & 6/7 \\
Solid Q1  & 7/7 & 7/7 & 6/7 & $3.57 \times 10^{-1}$ & 6/7 \\
Solid Q15 & 3/4 & 2/4 & 2/4 & $8.62 \times 10^{-1}$ & 3/4 \\
\addlinespace
\multicolumn{6}{l}{\textit{\textbf{Fluid Mechanics}}} \\
Fluid Q10 & 1/4 & 1/4 & 1/4 & $3.09 \times 10^{-2}$ & 1/4 \\
Fluid Q7  & 3/4 & 3/4 & 3/4 & $6.51 \times 10^{-1}$ & 3/4 \\
\bottomrule
\end{tabularx}
\end{table}

\begin{table}[!htbp]
\centering
\small
\caption{Test 2: Comparison of multi-step solution comparison evaluation process against a 0.001\% $L^2$ error threshold evaluated directly on the {framework-generated meshes}.}
\label{tab:thresholds_test2_updated}
\setlength{\tabcolsep}{6pt}
\renewcommand{\arraystretch}{1.3}

\begin{tabularx}{\linewidth}{l c >{\centering\arraybackslash}X c}
\toprule
\textbf{Validation case} & \textbf{\makecell{Pass count \\ (0.001\% Threshold)}} & \textbf{\makecell{Avg. Relative\\ $L^2$ norm (\%)\\ (Passed codes)}} & \textbf{\makecell{Multi-step\\solution\\comparison\\evaluation\\process}} \\
\midrule
\multicolumn{4}{l}{\textit{Solid Mechanics}} \\
Solid Q4  & 6/7 & $3.66 \times 10^{-11}$ & 6/7 \\
Solid Q1  & 6/7 & $1.14 \times 10^{-12}$ & 6/7 \\
Solid Q15 & 3/4 & $9.33 \times 10^{-11}$ & 3/4 \\
\addlinespace
\multicolumn{4}{l}{\textit{Fluid Mechanics}} \\
Fluid Q10 & 1/4 & $0$  & 1/4 \\
Fluid Q7  & 3/4 & $1.88 \times 10^{-4}$  & 3/4 \\
\bottomrule
\end{tabularx}
\end{table}

\FloatBarrier


\section{Mesh statistics for the Multi-agent framework} \label{app-mesh}
This appendix provides detailed mesh statistics for all benchmark problems executed in the multi-agent framework, so that the spatial discretizations used in the benchmark problems are verifiable. For each case, we report the number of elements, the maximum and minimum element size ($h_{\max}$, $h_{\min}$) and the number of degrees of freedom (DOFs). The tables are grouped by benchmark category to reflect differences in governing equations and discretization choices. These mesh statistics complement Sec.~\ref{subsec:results} by contextualizing performance variations across problems and by clarifying the discretization scale at which the agentic workflow successfully produced solutions.

\begin{table}
\centering
\scriptsize
\setlength{\tabcolsep}{4pt}
\renewcommand{\arraystretch}{1.12}
\caption{Mesh statistics for the solid mechanics benchmark set (Problems 1-16).}
\label{tab:mesh_solid}
\begin{tabular*}{\columnwidth}{@{\extracolsep{\fill}} l r r r r r}
\toprule
Case & Number of Elements & $h_{\max}$ (m) & $h_{\min}$ (m) & DOFs\\
\midrule
S1  & 320    & 0.05      & 0.05      & 370\\
S2  & 1,280  & 0.025     & 0.025     & 5,314\\
S3  & 45,679 & 0.00459   & 0.00233   & 184,378\\
S4  & 7,284  & 0.01147   & 0.00558   & 30,030\\
S5  & 46,831 & 0.00459   & 0.00229   & 188,820\\
S6  & 19,047 & 0.00717   & 0.00368   & 19,576\\
S7  & 5,120  & 0.0125    & 0.0125    & 5,314\\
S8  & 5,000  & 0.02      & 0.0107703 & 20,302\\
S9  & 8,000  & 0.01      & 0.01      & 32,482\\
S10 & 4,000  & 0.01414   & 0.01414   & 18,603\\
S11 & 4,826  & 0.01434   & 0.00563   & 22,612\\
S12 & 7,349  & 0.01147   & 0.00577   & 33,948\\
S13 & 4,781  & 0.01433   & 0.0073    & 22,317\\
S14 & 7,197  & 0.01147   & 0.00575   & 33,346\\
S15 & 14,160 & 0.00912   & 0.00422   & 58,172\\
S16 & 14,852 & 0.0023165 & 0.0012071 & 119,498\\
\bottomrule
\end{tabular*}
\end{table}

\begin{table}
\centering
\scriptsize
\setlength{\tabcolsep}{4pt}
\renewcommand{\arraystretch}{1.12}
\caption{Mesh statistics for the fluid mechanics benchmark set (Problems 1-15).}
\label{tab:mesh_fluid}
\begin{tabular*}{\columnwidth}{@{\extracolsep{\fill}} l r r r r r}
\toprule
Case & Number of Elements & $h_{\max}$ (m) & $h_{\min}$ (m) & DOFs\\
\midrule
F1  & 4,000  & 0.02     & 0.02     & 18,553       \\
F2  & 5,760  & 0.01667  & 0.01667  & 26,583       \\
F3  & 18,432 & 0.01473  & 0.01473  & 83,907    \\
F4  & 5,000  & 0.02     & 0.0107703& 20,302       \\
F5  & 65,536 & 0.007812 & 0.007812 & 296,195     \\
F6  & 27,600 & 0.105409 & 0.105409 & 125,653    \\
F7  & 7,246  & 0.025172 & 0.006954 & 33,567    \\
F8  & 16,384 & 0.007812 & 0.00625  & 74,531      \\
F9  & 16,000 & 0.01     & 0.01     & 81,324      \\
F10 & 8,192  & 0.022097 & 0.022097 & 41,732   \\
F11 & 19,200 & 0.016667 & 0.01     & 87,203       \\
F12 & 6,400  & 0.025    & 0.013463 & 29,303       \\
F13 & 4,000  & 0.01     & 0.01     & 2,111       \\
F14 & 19,956 & 0.954374 & 0.069473 & 90,782     \\
F15 & 4,096  & 0.03125  & 0.03125  & 18,441       \\
\bottomrule
\end{tabular*}
\end{table}

\begin{table}
\centering
\scriptsize
\setlength{\tabcolsep}{4pt}
\renewcommand{\arraystretch}{1.12}
\caption{Mesh statistics for the multiphysics benchmark set (Problems 1-8).}
\label{tab:mesh_multiphysics}
\begin{tabular*}{\columnwidth}{@{\extracolsep{\fill}} l r r r r r}
\toprule
Case & Number of Elements & $h_{\max}$ (m) & $h_{\min}$ (m) & DOFs \\
\midrule
M1 & 9,854  & 0.002     & 0.001022  & 5,056       \\
M2 & 80,000 & 0.007071  & 0.007071  & 40,401  \\
M3 & 32,768 & 0.011049  & 0.011049  & 32,772    \\
M4 & 4,096  & 0.098175  & 0.0625    & 29,059    \\
M5 & 19,200 & 0.00052202& 0.00052202& 174,806   \\
M6 & 4,208  & 0.035612  & 0.0070736 & 54,888    \\
M7 & 32,000 & 0.005     & 0.005     & 145,203     \\
M8 & 19,200 & 0.0083333 & 0.005     & 87,203      \\
\bottomrule
\end{tabular*}
\end{table}


\section{Computational cost}

We report computational cost for our agentic frameworks using three metrics: API calls, token usage (prompt, completion, and total), and wall-clock time in minutes~\cite{microsoft_prompt_tokens_2026,hf_tokenizers_pipeline}. Table ~\ref{tab:cost_comparison_new} reports these costs for runs using our fine-tuned GPT-OSS model on three representative example problems described in this section; this subset is included to give a practical sense of cost differences between the two-agent and multi-agent workflows, not to provide an exhaustive accounting over the full benchmark set. Wall-clock time includes both LLM latency and FEniCS runtime, including repeated executions during debugging, so it can vary substantially with hardware and system configuration and should be interpreted accordingly when comparing across different machines~\cite{mlperf_inference_rules,mlcommons_mlperf_inference_v4}.

\begin{table}[ht]
\centering
\small 
\caption{Computational cost comparison across frameworks (API calls, token usage, and wall-clock time).}
\label{tab:cost_comparison_new}
\setlength{\tabcolsep}{3pt} 
\renewcommand{\arraystretch}{1.3}

\begin{tabularx}{\linewidth}{l l c r r >{\centering\arraybackslash}X c}
\toprule
\textbf{Case} & \textbf{Framework} & \textbf{Calls} & \textbf{Prompt} & \textbf{Comp.} & \textbf{Total Tokens} & \textbf{\makecell{Wall-clock \\ (min)}} \\
\midrule

\hyperref[box:prob1]{Sample Problem 1} & Two-agent   & 16 & 36,128   & 9,648   & 45,776   & 6 \\
                   & Multi-agent & 47 & 632,083  & 84,353  & 716,436  & 11 \\
\midrule

\hyperref[box:prob2]{Sample Problem 2} & Two-agent   & 64 & 1,774,177 & 130,546 & 1,904,723 & 34 \\
                   & Multi-agent & 69 & 1,609,156 & 142,718 & 1,751,874 & 23 \\
\midrule

\hyperref[box:prob3]{Sample Problem 3} & Two-agent   & 101 & 3,070,132 & 140,884 & 3,211,016 & 42 \\
                   & Multi-agent & 70  & 1,909,111 & 188,303 & 2,097,414 & 29 \\

\bottomrule
 \end{tabularx}
\end{table}
\section{Sample Solution Plots} \label{appc}
\begin{figure} 
  \centering
  \includegraphics[width=\textwidth,keepaspectratio]{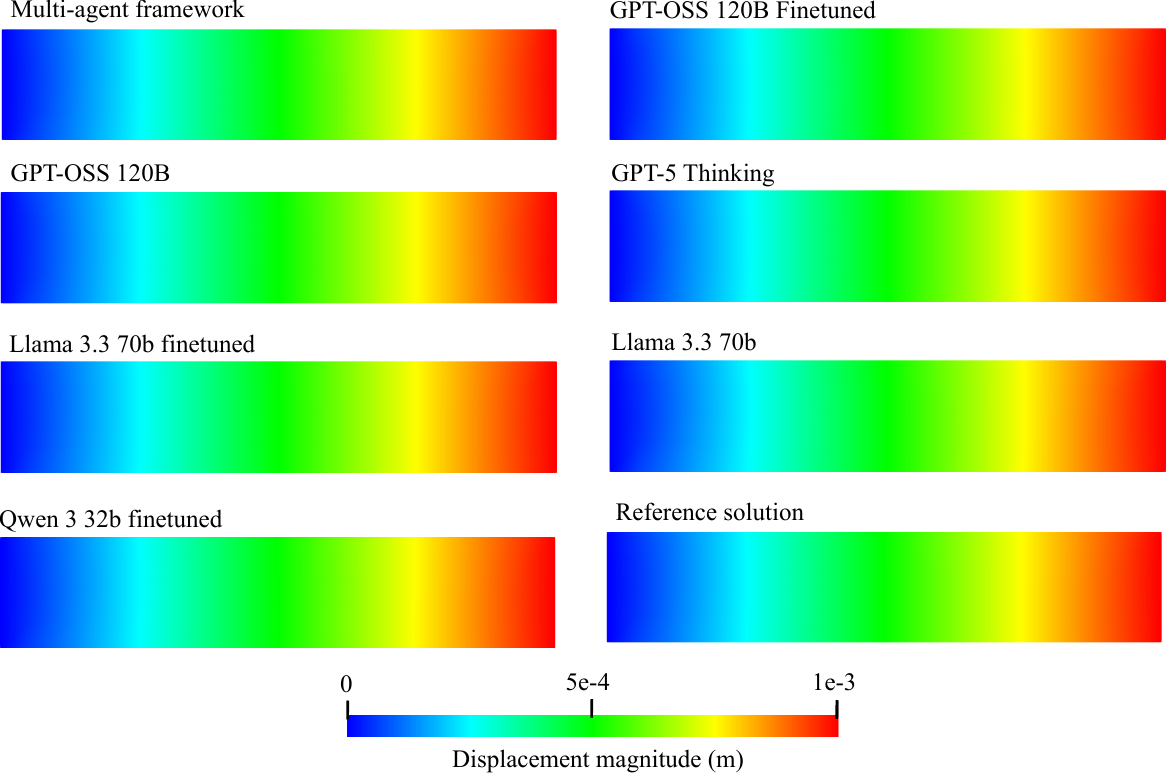}
\caption{Displacement magnitude fields for Solid Mechanics Q1 generated by the Two-agent frameworks, GPT-5 Thinking, and the Multi-agent framework.
}
  \label{fig: solid1}
\end{figure}

\begin{figure} 
  \centering
  \includegraphics[width=\textwidth,keepaspectratio]{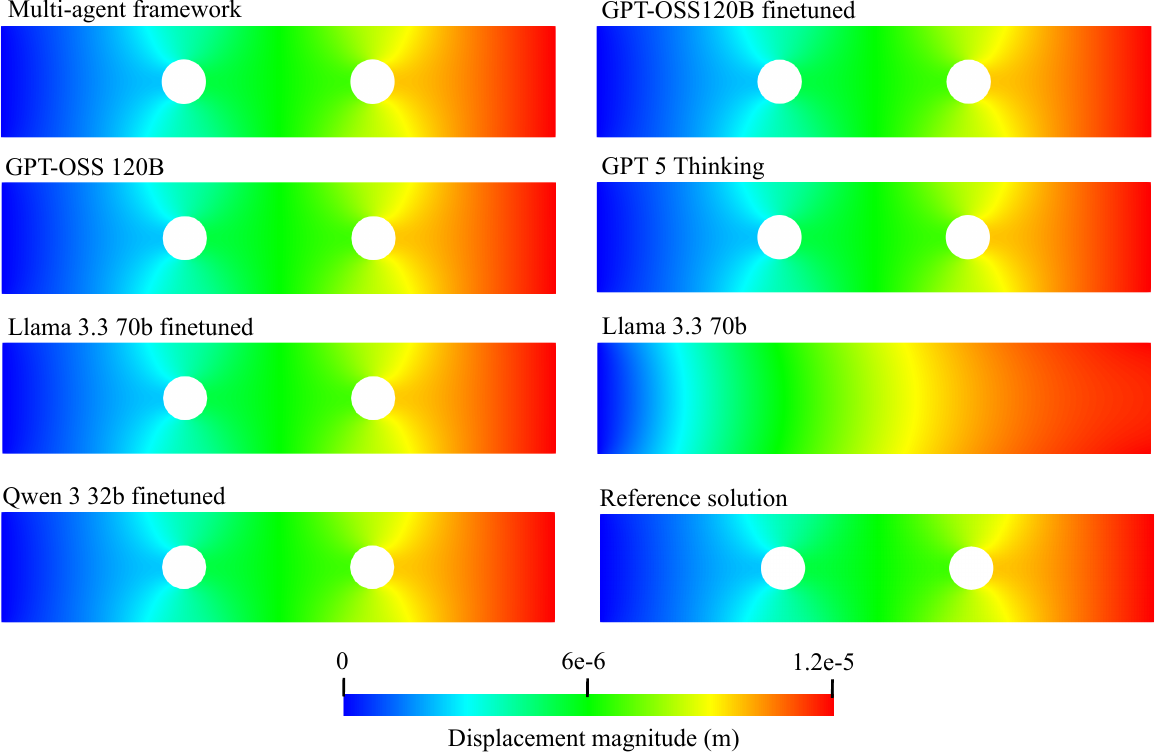}
\caption{ Solution fields for Solid Mechanics Q4 generated by the Two-agent frameworks, GPT-5 Thinking, and the Multi-agent framework. }
  \label{fig: solid4}
\end{figure}

\begin{figure} 
  \centering
  \includegraphics[width=\textwidth,keepaspectratio]{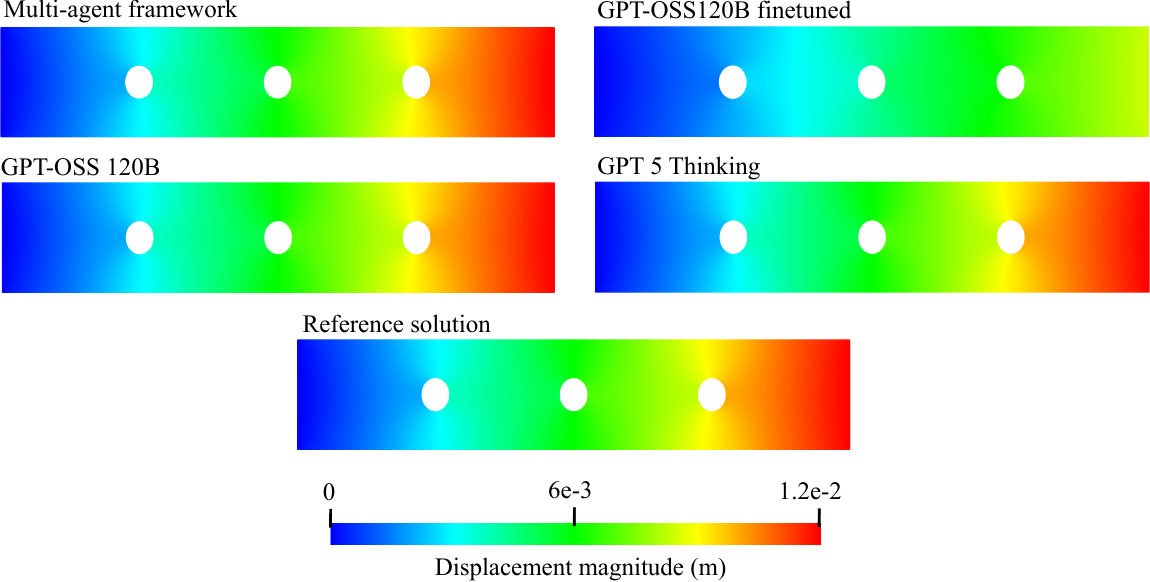}
\caption{Solution fields for Solid Mechanics Q15 generated by the Two-agent frameworks, GPT-5 Thinking, and the Multi-agent framework.}
  \label{fig: solid15}
\end{figure}

\begin{figure} 
  \centering
  \includegraphics[width=\textwidth,keepaspectratio]{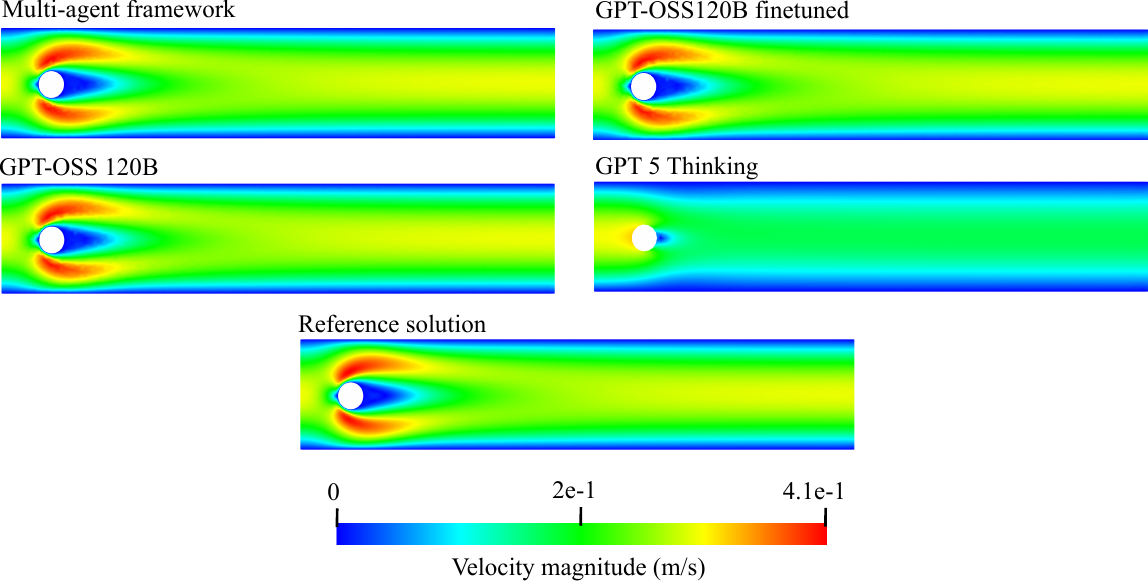}
\caption{Solution fields for Fluid Mechanics Q7 generated by the Two-agent frameworks, GPT-5 Thinking framework, and the Multi-agent framework.}
  \label{fig: fluid7}
\end{figure}

\begin{figure}[htbp] 
  \centering
  \includegraphics[width=.8\textwidth,  keepaspectratio]{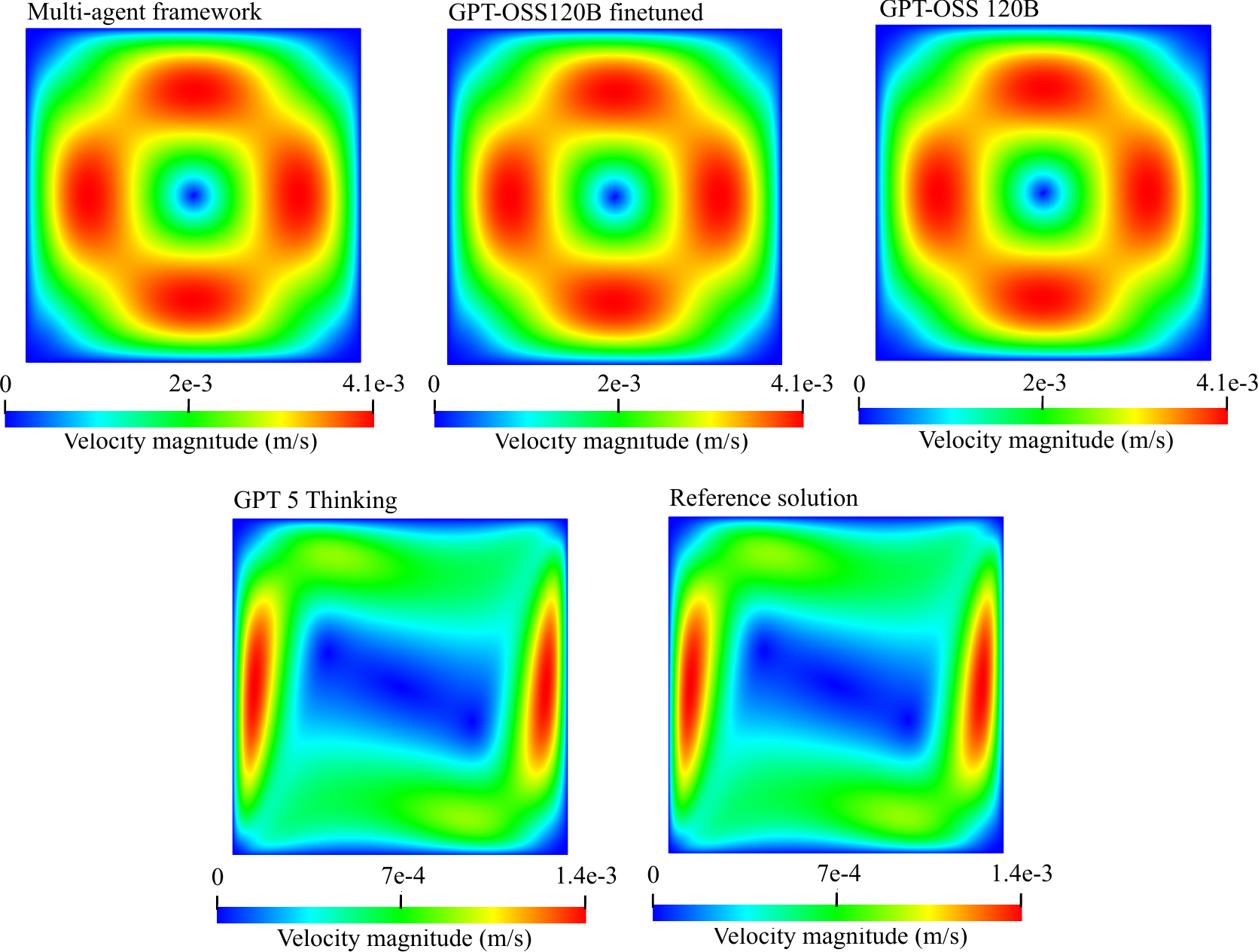}
  \vspace{10pt} 
\caption{Solution fields for Fluid Mechanics Q10 generated by the Two-agent frameworks, the GPT-5 Thinking framework, and the Multi-agent framework.}
  \label{fig: fluid10}
\end{figure}

In this section, we present representative solution plots for selected test-set problems to give the reader a visual sense of the fields produced by our agentic frameworks and GPT-5 Thinking, alongside the corresponding reference solution plots. Figure~\ref{fig: solid1} corresponds to Solid Q1 (Easy) and shows the displacement magnitude over a rectangular plate. All frameworks produce displacement fields that are visually identical to the reference solution. Figure~\ref{fig: solid4} corresponds to Solid Q4 (Easy) and shows the displacement magnitude over a rectangular plate with two circular holes. Most frameworks produce outputs with displacement fields that match the reference pattern, but Llama 3.3 70B in two-agent framework produces a different result because the circular holes were missed in the geometry. Figure~\ref{fig: solid15} corresponds to Solid Q15 (Medium) and shows the displacement magnitude over a rectangular plate with three circular holes. The Multi-agent framework, GPT-OSS 120B and GPT-5 Thinking produce plots with displacement fields that are consistent with the reference solution. However, GPT-OSS 120B Finetuned, which yields an executable output, produces a different displacement magnitude. Figure~\ref{fig: fluid7} corresponds to Fluid Q7 (Medium) and shows the velocity magnitude field in a rectangular channel with a circular obstacle.  The Multi-agent framework and GPT-OSS 120B variants (base and finetuned) produce velocity magnitude fields that align closely with the reference plot, including similar high-velocity regions around the obstacle and downstream structure. In contrast, the GPT-5 Thinking output has noticeably weaker spatial variation around and downstream of the obstacle. Figure~\ref{fig: fluid10} corresponds to Fluid Q10 (Medium) and shows the velocity magnitude field in a unit-square cavity. GPT-5 Thinking matches the reference solution in the overall field structure, while the multi-agent framework and the GPT-OSS 120B variants produce identical fields to each other but with a different overall spatial pattern than the GPT-5 Thinking and the reference solution.

\FloatBarrier

\section{Benchmark Problems}
\label{subsec:benchmark}
Here, we provide the complete list of 39 computational mechanics problems that we used for the evaluation of our agentic systems. For better readability, the problems are shown here in LaTeX fonts, but they were prompted to the agentic systems as plain text with Unicode symbols.

\paragraph{Fluid Mechanics Problems}
\vspace{.25cm}

\begin{tcolorbox}[enhanced,breakable,colback=gray!10,colframe=gray!50,title={Fluid Mechanics Problem 1 (Easy)}, label=fm_q1, 
    phantom={\phantomsection\label{box:fm_q1}}]

\noindent \textbf{Geometry:} \\
Let $\Omega = (0, L) \times (0, H)$ be a rectangular channel with length $L = 2.0$~m and height $H = 0.20$~m.
Boundary partition: $\Gamma_{\text{in}} = \{0\}\times[0, H]$, $\Gamma_{\text{out}} = \{L\}\times[0, H]$, $\Gamma_w = (0, L)\times\{0\} \cup (0, L)\times\{H\}$.

\medskip
\noindent \textbf{Mesh:} \\
Use a uniform structured mesh of $100 \times 10$ elements.

\medskip
\noindent \textbf{Model:} \\
Steady incompressible Stokes flow in $\Omega$. Let velocity be $u = (u_x, u_y)$, pressure $p$ and Cauchy stress tensor $\sigma$.

\medskip
\noindent \textbf{Parameters:} \\
Dynamic viscosity $\mu = 1.0$~Pa$\cdot$s; density $\rho = 1.0$~kg$\cdot$m$^{-3}$.

\medskip
\noindent \textbf{Boundary conditions:} \\
Walls ($\Gamma_w$): no-slip and no penetration. \\
Inlet ($\Gamma_{\text{in}}$): traction (natural) boundary, $\sigma(u,p) n = - p_{\text{in}} n$ on $\Gamma_{\text{in}}$, with $p_{\text{in}} = 1.0$~Pa. \\
Outlet ($\Gamma_{\text{out}}$): traction (natural) boundary, $\sigma(u,p) n = - p_{\text{out}} n$ on $\Gamma_{\text{out}}$, with $p_{\text{out}} = 0$~Pa. \\
$n$ denotes the unit outward normal vector.

\medskip
\noindent \textbf{Output:} \\
Save a color map of the speed $|u|$ over $\Omega$ to \texttt{q1\_speed.png}. \\
Save the velocity field ($u$) and pressure field ($p$) to \texttt{q1\_soln.xdmf}.

\end{tcolorbox}

\begin{tcolorbox}[enhanced,breakable,colback=gray!10,colframe=gray!50,title={Fluid Mechanics Problem 2 (Easy)}, label=fm_q2, 
    phantom={\phantomsection\label{box:fm_q2}}]

\noindent \textbf{Geometry:} \\
Let $\Omega = (0, L) \times (0, H)$ be a rectangular channel with length $L = 2.0$~m and height $H = 0.20$~m.
Boundary partition: $\Gamma_{\text{in}} = \{0\}\times[0, H]$, $\Gamma_{\text{out}} = \{L\}\times[0, H]$, $\Gamma_w = (0, L)\times\{0\} \cup (0, L)\times\{H\}$.

\medskip
\noindent \textbf{Mesh:} \\
Use a uniform structured mesh of $120 \times 12$ elements.

\medskip
\noindent \textbf{Model:} \\
Steady incompressible Stokes flow with body force for velocity $u = (u_x, u_y)$ and pressure $p$ in $\Omega$. \\
Uniform body force: $f = (1.0, 0.0)$~N~m$^{-3}$.

\medskip
\noindent \textbf{Parameters:} \\
Dynamic viscosity $\mu = 1.0$~Pa$\cdot$s; density $\rho = 1.0$~kg~m$^{-3}$.

\medskip
\noindent \textbf{Boundary conditions:} \\
Walls ($\Gamma_w$): no-slip and no penetration. \\
Inlet and outlet ($\Gamma_{\text{in}} \cup \Gamma_{\text{out}}$): traction-free natural condition: $(-p I + \mu(\nabla u + \nabla u^T)) n = 0$. \\
where $n$ denotes the unit outward normal vector.

\medskip
\noindent \textbf{Output:} \\
Save a color map of speed $|u|$ over $\Omega$ as \texttt{q2\_speed.png}. \\
Also, save the velocity field ($u$) and pressure field ($p$) to \texttt{q2\_solution.xdmf}.
\end{tcolorbox}

\begin{tcolorbox}[enhanced,breakable,colback=gray!10,colframe=gray!50,title={Fluid Mechanics Problem 3 (Easy)}, label=fm_q3, 
    phantom={\phantomsection\label{box:fm_q3}}]

\noindent \textbf{Geometry:} \\
Let $\Omega = (0, 1) \times (0, 1)$ be a unit square cavity.
Boundary partition: $\Gamma_{\text{left}} = \{0\}\times[0, 1]$, $\Gamma_{\text{right}} = \{1\}\times[0, 1]$, $\Gamma_{\text{bottom}} = (0, 1)\times\{0\}$, $\Gamma_{\text{top}} = (0, 1)\times\{1\}$.

\medskip
\noindent \textbf{Mesh:} \\
Use a uniform structured mesh of $96 \times 96$ elements.

\medskip
\noindent \textbf{Model:} \\
Lid driven cavity flow for steady incompressible Stokes flow for velocity $u = (u_x, u_y)$ and pressure $p$ in $\Omega$. Use an inf-sup stable finite element pair (Taylor-Hood P2-P1).

\medskip
\noindent \textbf{Parameters:} \\
Density $\rho = 1.0$~kg~m$^{-3}$; dynamic viscosity $\mu = 1.0$~Pa$\cdot$s.

\medskip
\noindent \textbf{Boundary conditions:} \\
Lid ($\Gamma_{\text{top}}$): prescribed tangential lid motion, $u = (1, 0)$. \\
Other walls ($\Gamma_{\text{left}} \cup \Gamma_{\text{right}} \cup \Gamma_{\text{bottom}}$): no-slip and no penetration, $u = (0, 0)$.

\medskip
\noindent \textbf{Output:} \\
Save a color map of speed $|u|$ over $\Omega$ as \texttt{q3\_speed.png}. \\
Also, save the velocity field ($u$) and pressure field ($p$) to \texttt{q3\_soln.xdmf}.
\end{tcolorbox}

\begin{tcolorbox}[enhanced,breakable,colback=gray!10,colframe=gray!50,title={Fluid Mechanics Problem 4 (Easy)}, label=fm_q4, 
    phantom={\phantomsection\label{box:fm_q4}}]

\noindent \textbf{Geometry:} \\
Let $\Omega = (0, L) \times (0, H)$ be a rectangular channel with length $L = 2.0$~m and height $H = 0.20$~m.
Boundary partition: $\Gamma_{\text{in}} = \{0\}\times[0, H]$, $\Gamma_{\text{out}} = \{L\}\times[0, H]$, $\Gamma_w = (0, L)\times\{0\} \cup (0, L)\times\{H\}$.

\medskip
\noindent \textbf{Mesh:} \\
Use a uniform structured mesh of $160 \times 16$ elements.

\medskip
\noindent \textbf{Model:} \\
Steady incompressible Navier-Stokes for velocity $u = (u_x, u_y)$ and pressure $p$ in $\Omega$.

\medskip
\noindent \textbf{Parameters:} \\
Dynamic viscosity $\mu = 0.01$~Pa~s and $\rho = 1$~kg~m$^{-3}$.

\medskip
\noindent \textbf{Boundary conditions:} \\
Inlet ($\Gamma_{\text{in}}$): $u_x(y) = 6 \bar{U} (y/H) (1 - y/H)$, $u_y = 0$, with mean velocity $\bar{U} = 2.5$~m~s$^{-1}$. \\
Walls ($\Gamma_w$): no-slip and no penetration. \\
Outlet ($\Gamma_{\text{out}}$): traction-free: $(-p I + \mu (\nabla u + \nabla u^T)) n = 0$. \\
where $n$ is outward normal unit vector.

\medskip
\noindent \textbf{Output:} \\
Save a color map of $u_x$ over $\Omega$ as \texttt{q4\_ux.png}. \\
Also, save the velocity field ($u$) and pressure field ($p$) to \texttt{q4\_soln.xdmf}.
\end{tcolorbox}

\begin{tcolorbox}[enhanced,breakable,colback=gray!10,colframe=gray!50,title={Fluid Mechanics Problem 5 (Easy)}, label=fm_q5, 
    phantom={\phantomsection\label{box:fm_q5}}]

\noindent \textbf{Geometry:} \\
Let $\Omega = (0, 1) \times (0, 1)$ be a unit square cavity.
Boundary partition: $\Gamma_{\text{left}} = \{0\}\times[0, 1]$, $\Gamma_{\text{right}} = \{1\}\times[0, 1]$, $\Gamma_{\text{bottom}} = (0, 1)\times\{0\}$, $\Gamma_{\text{top}} = (0, 1)\times\{1\}$.

\medskip
\noindent \textbf{Mesh:} \\
Use a uniform structured mesh of $128 \times 128$ elements.

\medskip
\noindent \textbf{Model:} \\
Lid driven cavity flow for steady incompressible Navier-Stokes for velocity $u = (u_x, u_y)$ and pressure $p$ in $\Omega$.

\medskip
\noindent \textbf{Parameters:} \\
Given density $\rho = 1$~kg~m$^{-3}$, Dynamic viscosity $\mu = 0.01$~Pa$\cdot$s.

\medskip
\noindent \textbf{Boundary conditions:} \\
Lid ($\Gamma_{\text{top}}$): $u = (1, 0)$. \\
Walls ($\Gamma_{\text{left}} \cup \Gamma_{\text{right}} \cup \Gamma_{\text{bottom}}$): no-slip and no penetration. \\
Pressure gauge for uniqueness: fix $p = 0$ at $[0,0]$.

\medskip
\noindent \textbf{Output:} \\
Save the color map of speed $|u|$ over $\Omega$ as \texttt{q5\_speed.png}. \\
Also, save the velocity field ($u$) and pressure field ($p$) to \texttt{q5\_soln.xdmf}.
\end{tcolorbox}

\begin{tcolorbox}[enhanced,breakable,colback=gray!10,colframe=gray!50,title={Fluid Mechanics Problem 6 (Medium)}, label=fm_q6, 
    phantom={\phantomsection\label{box:fm_q6}}]

\noindent \textbf{Geometry:} \\
Let $H = 1.0$~m. The domain $\Omega$ is a 2D channel with a backward-facing step at $x = 0$. It is defined as the union of two rectangular regions: \\
Upstream channel: $\Omega_1 = \{ (x, y) : -3H \le x \le 0, \ 0 \le y \le H \}$ \\
Downstream channel: $\Omega_2 = \{ (x, y) : 0 < x \le 20H, \ 0 \le y \le 2H \}$ \\
The total domain $\Omega = \Omega_1 \cup \Omega_2$ has a 2:1 expansion ratio.

\medskip
\noindent \textbf{Model:} \\
Steady, incompressible Navier-Stokes equations for velocity $u = (u_x, u_y)$ and pressure $p$ in $\Omega$.

\medskip
\noindent \textbf{Boundary conditions:} \\
Inlet (at $x = -3H$, for $y \in [0, H]$): Prescribed velocity: $u_x(y) = 6 \bar{U} (y/H) (1 - y/H)$, $u_y = 0$. \\
Solid Walls (No-slip and no penetration): $u_x = 0, u_y = 0$ on the following boundaries: \\
Bottom Wall: $\{ (x, y) : y = 0, \ -3H \le x \le 20H \}$ \\
Top Wall: $\{ (x, y) : y = H, \ -3H \le x \le 0 \} \cup \{ (x, y) : y = 2H, \ 0 \le x \le 20H \}$ \\
Step Wall: $\{ (x, y) : x = 0, \ H \le y \le 2H \}$ \\
Outlet (at $x = 20H$, for $y \in [0, 2H]$): Traction-free outflow: $(-p I + \mu(\nabla u + \nabla u^T)) n = 0$.

\medskip
\noindent \textbf{Parameters:} \\
Density $\rho = 1$~kg~m$^{-3}$; Dynamic viscosity $\mu = 0.01$~Pa$\cdot$s; Mean inlet speed $\bar{U} = 1.0$~m~s$^{-1}$.

\medskip
\noindent \textbf{Output:} \\
Compute the wall shear stress ($\tau_w$) on the top wall for $x \in [0, 20H]$: $\tau_w(x,H) = \mu (\partial u_x / \partial y)$. \\
The wall shear stress is used to find the re-attachment point, where $\tau_w(x,H) = 0$. \\
Save velocity ($u$) field as \texttt{q6\_u.png}. \\
Also, save the velocity ($u$) and pressure ($p$) solution fields in XDMF format as \texttt{q6\_soln.xdmf}.
\end{tcolorbox}

\begin{tcolorbox}[enhanced,breakable,colback=gray!10,colframe=gray!50,title={Fluid Mechanics Problem 7 (Medium)}, label=fm_q7, 
    phantom={\phantomsection\label{box:fm_q7}}]

\noindent \textbf{Geometry:} \\
Let $\Omega = [0, 2.2] \times [0, 0.41]$~m be a rectangular channel with a circular hole of radius $R = 0.05$~m centered at $(0.20, 0.20)$~m. 

\medskip
\noindent \textbf{Model:} \\
Steady incompressible Navier-Stokes for velocity $u = (u_x, u_y)$ and pressure $p$ in $\Omega$.

\medskip
\noindent \textbf{Boundary conditions:} \\
Inlet ($x = 0$): $u_x(y) = 6 \bar{U} y (H - y) / H^2$, $u_y = 0$, where $H = 0.41$. \\
Walls ($y = 0$ and $y = 0.41$) and circular boundary: no-slip and no penetration. \\
Outlet ($x = 2.2$): traction-free.

\medskip
\noindent \textbf{Parameters:} \\
Dynamic viscosity $\mu = 0.001$~Pa~s, Density $\rho = 1$~kg~m$^{-3}$, and Mean inlet velocity $\bar{U} = 0.2$~m/s.

\medskip
\noindent \textbf{Output:} \\
Drag coefficient $C_D$: \\
Compute $C_D$ from the drag force on the circle $F_D$ via $C_D = 2 F_D / (\rho \bar{U}^2 D)$. \\
Save a color map of speed $|u|$ over $\Omega$ as \texttt{q7\_speed.png}. \\
Also, save the velocity field ($u$) and pressure field ($p$) to \texttt{q7\_soln.xdmf}.
\end{tcolorbox}

\begin{tcolorbox}[enhanced,breakable,colback=gray!10,colframe=gray!50,title={Fluid Mechanics Problem 8 (Easy)}, label=fm_q8, 
    phantom={\phantomsection\label{box:fm_q8}}]

\noindent \textbf{Geometry:} \\
Let $\Omega = [0, 1] \times [0, 0.20]$~m be a rectangular channel.
Boundary partition: $\Gamma_{y0} = (0, 1)\times\{0\}$, $\Gamma_{yH} = (0, 1)\times\{0.20\}$.

\medskip
\noindent \textbf{Mesh:} \\
Use a uniform structured mesh of $128 \times 32$ elements.

\medskip
\noindent \textbf{Model:} \\
Steady incompressible Navier-Stokes with body force for velocity $u = (u_x, u_y)$ and pressure $p$ in $\Omega$. \\
Drive with a uniform body force $f = (G, 0)$.

\medskip
\noindent \textbf{Boundary conditions:} \\
Periodic in $x$: $u$ and $p$ are periodic in $x$ direction. \\
Walls ($\Gamma_{y0}$ and $\Gamma_{yH}$): no-slip and no penetration. \\
Pressure gauge: pin $p = 0$ at $[0,0]$.

\medskip
\noindent \textbf{Parameters:} \\
Density $\rho = 1$~kg~m$^{-3}$. Dynamic viscosity $\mu = 0.01$~Pa~s. $G = 1$~N~m$^{-3}$.

\medskip
\noindent \textbf{Output:} \\
Save the velocity field ($u$) and pressure field ($p$) to \texttt{q9\_soln.xdmf}.
\end{tcolorbox}

\begin{tcolorbox}[enhanced,breakable,colback=gray!10,colframe=gray!50,title={Fluid Mechanics Problem 9 (Medium)}, label=fm_q9, 
    phantom={\phantomsection\label{box:fm_q9}}]
    
\noindent \textbf{Geometry:} \\
Let $\Omega = [0, 2.0] \times [0, 0.20]$~m be a rectangular channel.

\medskip
\noindent \textbf{Mesh:} \\
Use a uniform mesh composed of $200 \times 20$ elements.

\medskip
\noindent \textbf{Model:} \\
Fluid flow in the domain is governed by steady incompressible Navier-Stokes equations with velocity $u = (u_x, u_y)$ and pressure $p$. \\
After solving the fluid flow problem, solve the steady advection-diffusion for the transport of a chemical with concentration $c$, where the diffusion coefficient is $\kappa$ and the advective transport is driven by the velocity $u$.

\medskip
\noindent \textbf{Boundary conditions:} \\
Flow ($u, p$): \\
Inlet ($x = 0$): $u_x(y) = 6 \bar{U} (y/H) (1 - y/H)$, with Mean inflow speed $\bar{U} = 0.1$~m~s$^{-1}$ and $H = 0.20$~m, $u_y = 0$. \\
Walls ($y = 0$ and $y = 0.20$): no-slip and no penetration. \\
Outlet ($x = 2.0$): traction-free.\\
Concentration ($c$): \\
Inlet ($x = 0$): Dirichlet $c = 0$. \\
Walls ($y = 0$ and $y = 0.20$): impermeable, $-\kappa \nabla c \cdot n = 0$, where $n$ is unit normal vector. \\
Outlet ($x = 2.0$): Dirichlet $c = 1$.

\medskip
\noindent \textbf{Parameters:} \\
Density $\rho = 1$~kg~m$^{-3}$; Dynamic viscosity $\mu = 0.01$~Pa~s. \\
Diffusivity $\kappa = 1.0 \times 10^{-3}$~m$^2$~s$^{-1}$.

\medskip
\noindent \textbf{Output:} \\
Save the concentration field as a color map \texttt{q10\_conc.png}. \\
Also, save the velocity ($u$), pressure ($p$), and concentration ($c$) fields to \texttt{q10\_solution.xdmf}.
\end{tcolorbox}

\begin{tcolorbox}[enhanced,breakable,colback=gray!10,colframe=gray!50,title={Fluid Mechanics Problem 10 (Medium)}, label=fm_q10, 
    phantom={\phantomsection\label{box:fm_q10}}]

\noindent \textbf{Geometry:} \\
Let $\Omega = [0, 1] \times [0, 1]$~m be a unit square cavity.

\medskip
\noindent \textbf{Model (Boussinesq, steady):} \\
Solve the steady incompressible Navier-Stokes with body force $f$ coupled to a steady advection-diffusion equation for temperature $T$. \\
The notation for velocity is $u = (u_x, u_y)$, pressure is $p$, and temperature is $T$. \\
The body force $f$ is modeled as a Boussinesq buoyancy term, $f = [0, \rho g \beta (T - T_{\text{ref}})]$, where $\rho$ is fluid density, $g$ is gravitational acceleration, $\beta$ is volumetric thermal expansion coefficient, $T_{\text{ref}}$ is reference temperature.

\medskip
\noindent \textbf{Boundary conditions:} \\
Left wall ($x = 0$): $T = 1$ (hot), no-slip and no penetration. \\
Right wall ($x = 1$): $T = 0$ (cold), no-slip and no penetration. \\
Top/bottom ($y = 1$ and $y = 0$): adiabatic $\partial T/\partial n = 0$, no-slip and no penetration. \\
where $n$ is the outward unit normal vector.

\medskip
\noindent \textbf{Parameters:} \\
Density $\rho = 1$~kg/m$^3$. \\
Dynamic viscosity ($\mu$): $1.5 \times 10^{-5}$~Pa$\cdot$s. \\
Thermal diffusivity $\alpha = 2.1 \times 10^{-5}$~m$^2$~s$^{-1}$. \\
$g \beta = 3.15 \times 10^{-5}$~m~s$^{-2}$~K$^{-1}$. \\
$T_{\text{ref}} = 0.5$~K.

\medskip
\noindent \textbf{Output:} \\
Temperature field: save a color map as \texttt{q11\_T.png}. \\
Report the average Nusselt number at the left wall. \\
Also, save the velocity field ($u$), pressure field ($p$), and temperature field ($T$) to \texttt{q11\_solution.xdmf}.
\end{tcolorbox}

\begin{tcolorbox}[enhanced,breakable,colback=gray!10,colframe=gray!50,title={Fluid Mechanics Problem 11 (Medium)}, label=fm_q11, 
    phantom={\phantomsection\label{box:fm_q11}}]

\noindent \textbf{Geometry:} \\
Let $\Omega = [0, 2.0] \times [0, 0.20]$~m be a 2D rectangular channel (Length $L = 2.0$~m, Height $H = 0.20$~m).

\medskip
\noindent \textbf{Mesh:} \\
Structured mesh with $240 \times 24$ subdivisions.

\medskip
\noindent \textbf{Model:} \\
Steady, incompressible, non-Newtonian flow. \\
The governing equations for velocity $u = (u_x, u_y)$, and pressure $p$ are: \\
$\rho (u \cdot \nabla)u = -\nabla p + \nabla \cdot \tau$ \\
$\nabla \cdot u = 0$ \\
The stress tensor $\tau$ is defined by a power-law model: $\tau = 2 \mu_{\text{eff}} D$, where $D = (\nabla u + (\nabla u)^T)/2$ is the strain-rate tensor and $\mu_{\text{eff}}$ is the effective viscosity.

\medskip
\noindent \textbf{Material (Power-Law Fluid):} \\
The effective viscosity $\mu_{\text{eff}}$ is a function of the shear rate $|D| = (2 D:D)^{1/2} + 10^{-8}$. \\
$\mu_{\text{eff}}(|D|) = \mu_0 [|D|^{n-1}]$. \\
Density $\rho = 1.0$~kg~m$^{-3}$. \\
Consistency index $\mu_0 = 0.5$~Pa$\cdot$s$^n$. \\
Flow behavior index $n = 0.5$.

\medskip
\noindent \textbf{Boundary conditions:} \\
Inlet ($x = 0$): $u_x(y) = 6 \bar{U} y (H - y) / H^2$, with $\bar{U} = 1.0$~m~s$^{-1}$. $u_y = 0$. \\
Walls ($y = 0$ and $y = H$): No-slip and no penetration. \\
Outlet ($x = 2.0$): traction-free, $(-pI + \tau) n = 0$.

\medskip
\noindent \textbf{Output:} \\
Save a color map of the velocity magnitude $|u|$ as \texttt{q12\_speed.png}. \\
Extract the streamwise velocity profile $u_x(y)$ at mid-channel ($x = 1.0$) and save it as \texttt{q12\_profile.csv}. \\
Also, save the velocity field ($u$), pressure field ($p$), and effective viscosity ($\mu_{\text{eff}}$) to \texttt{q12\_solution.xdmf}. \\
Report the maximum velocity $u_x(y)$ at $x = L/2$.
\end{tcolorbox}

\begin{tcolorbox}[enhanced,breakable,colback=gray!10,colframe=gray!50,title={Fluid Mechanics Problem 12 (Hard)}, label=fm_q12, 
    phantom={\phantomsection\label{box:fm_q12}}]

\noindent \textbf{Geometry:} \\
Let $\Omega = [0, 2.0] \times [0, 0.20]$~m be a 2-D rectangular channel of length $L = 2.0$~m and height $H = 0.20$~m.

\medskip
\noindent \textbf{Model:} \\
Steady incompressible Navier-Stokes for velocity $u = (u_x, u_y)$ and pressure $p$ in $\Omega$, coupled to a steady advection-diffusion equation for temperature $T$: \\
Momentum: $\rho (u \cdot \nabla) u - \nabla \cdot [2 \mu(T) \varepsilon(u)] + \nabla p = 0$, where $\varepsilon(u) = (\nabla u + (\nabla u)^T)/2$. \\
Mass conservation: $\nabla \cdot u = 0$. \\
Energy: $u \cdot \nabla T - \kappa \nabla^2 T = 0$. \\
Temperature-dependent viscosity: $\mu(T) = \mu_{\text{ref}}  e^{-\beta (T - T_{\text{ref}})}$.

\medskip
\noindent \textbf{Boundary conditions:} \\
Inlet ($x = 0$): $u_x(y) = 6 \bar{U} y (H - y) / H^2$, $u_y = 0$; $T = T_{\text{ref}}$. \\
Walls ($y = 0$ and $y = H$): no-slip and no penetration for the flow. \\
The boundary conditions at the walls for the temperature equation are: \\
Bottom wall ($y = 0$): Dirichlet $T = T_{\text{ref}} + 10$~K. \\
Top wall ($y = H$): $\partial T/\partial n = 0$. \\
Outlet ($x = L = 2.0$): traction-free for the flow. \\
For temperature: $-\kappa \partial T/\partial n = 0$. \\
where $n$ is outward normal vector.

\medskip
\noindent \textbf{Parameters:} \\
Density: $\rho = 1.0$~kg~m$^{-3}$. \\
Mean inlet speed: $\bar{U} = 1.0$~m~s$^{-1}$. \\
Reference viscosity: $\mu_{\text{ref}} = 0.02$~Pa$\cdot$s. \\
$\beta = 0.05$~K$^{-1}$. \\
Reference temperature: $T_{\text{ref}} = 300$~K. \\
Thermal diffusivity: $\kappa = 1.0\times 10^{-3}$~m$^2$~s$^{-1}$.

\medskip
\noindent \textbf{Output:} \\
Save $\mu(x, y)$ as a color map image \texttt{q13\_mu.png}. \\
Extract the streamwise velocity profile $u_x(y)$ along the mid-length line ($x = 1.0$) and save it to a CSV file named \texttt{q13\_profile.csv} with columns: y, ux. \\
Export solution fields ($u, p, T, \mu$) in XDMF format for post-processing as \texttt{q13\_solution.xdmf}.
\end{tcolorbox}

\begin{tcolorbox}[enhanced,breakable,colback=gray!10,colframe=gray!50,title={Fluid Mechanics Problem 13 (Medium)}, label=fm_q13, 
    phantom={\phantomsection\label{box:fm_q13}}]

\noindent \textbf{Geometry:} \\
Solve for the transport of a chemical in a rectangular channel $\Omega = (0, L) \times (0, H)$ with length $L = 1.0$~m and height $H = 0.10$~m.
Boundary partition: $\Gamma_{\text{in}} = \{0\}\times[0, H]$, $\Gamma_{\text{out}} = \{L\}\times[0, H]$, $\Gamma_w = (0, L)\times\{0\} \cup (0, L)\times\{H\}$.

\medskip
\noindent \textbf{Mesh:} \\
Use a uniform mesh composed of $100 \times 10$ elements.

\medskip
\noindent \textbf{Model:} \\
Steady advection-diffusion for a scalar concentration $c$ in $\Omega$ with a given velocity field $u(x,y) = (u_x(y), 0)$, \; $u_x(y) = U_{\text{max}} \cdot [4 y (H - y) / H^2]$.

\medskip
\noindent \textbf{Boundary conditions:} \\
Inlet ($\Gamma_{\text{in}}$): $c = 0$. \\
Outlet ($\Gamma_{\text{out}}$): $c = 1$. \\
Walls ($\Gamma_w$): zero diffusive normal flux.

\medskip
\noindent \textbf{Parameters:} \\
$L = 1.0$~m; $H = 0.10$~m; $U_{\text{max}} = 0.75$~m~s$^{-1}$. \\
Diffusivity of the chemical: $D = 1.0\times10^{-5}$~m$^2$~s$^{-1}$.

\medskip
\noindent \textbf{Output:} \\
Save the resulting concentration field output in xdmf format.

\medskip
\noindent \textbf{Hint:} \\
Check the Peclet number (using the mesh size as length scale) to determine whether a stabilization scheme is required. If stabilization is required, include it in your formulation.
\end{tcolorbox}

\begin{tcolorbox}[enhanced,breakable,colback=gray!10,colframe=gray!50,title={Fluid Mechanics Problem 14 (Hard)}, label=fm_q14, 
    phantom={\phantomsection\label{box:fm_q14}}]

\noindent \textbf{Geometry:} \\
Solve the turbulent flow over a cylinder in the square domain $\Omega = [-30 D, +30 D] \times [-30 D, +30 D]$. The circular cylinder of diameter $D$ is placed at $(0, 0)$. 

\medskip
\noindent \textbf{Model:} \\
Solve the incompressible, unsteady Navier-Stokes equations in $\Omega$ using a Variational Multiscale (residual-based VMS) formulation with streamline-upwind, pressure-stabilizing Petrov-Galerkin, and grad-div stabilizations. \\
We use $u$ to denote the velocity and $p$ to denote the pressure.

\medskip
\noindent \textbf{Boundary conditions:} \\
Impose a uniform inflow $u = (U, 0)$ at $x = -30 D$. Impose traction-free outflow with reference pressure $p = 0$ at $x = +30 D$. \\
Impose no slip and no penetration on the top and bottom boundaries at $y = \pm 30 D$, and on the cylinder surface. Use the initial condition $u = (0,0)$ and $p = 0$, with an optional small perturbation to trigger vortex shedding.

\medskip
\noindent \textbf{Parameters:} \\
Set $U = 1.0$~m/s, kinematic viscosity $\nu = 2.56 \times 10^{-5}$~m$^2$/s, and density $\rho = 1.0$~kg/m$^3$, diameter $D = 1$~m.

\medskip
\noindent \textbf{Output:} \\
Report the mean drag coefficient computed over $t \in [8.0 \text{ s}, 10.0 \text{ s}]$. \\
Save the final velocity field ($u$) and pressure field ($p$) at $t=10.0$~s to \texttt{vms\_solution.xdmf}.
\end{tcolorbox}

\begin{tcolorbox}[enhanced,breakable,colback=gray!10,colframe=gray!50,title={Fluid Mechanics Problem 15 (Medium)}, label=fm_q15, 
    phantom={\phantomsection\label{box:fm_q15}}]

\noindent \textbf{Geometry:} \\
Analyze the fluid flow in the unit square $[0, 1] \times [0, 1]$.

\medskip
\noindent \textbf{Model:} \\
Solve the unsteady incompressible Navier-Stokes equations for velocity $U = (u, v)$ and pressure $p$ in the domain.

\medskip
\noindent \textbf{Boundary conditions:} \\
Impose periodic boundary conditions on all four sides. \\
The initial condition for the velocity is: \\
$u(x, y, 0) = \sin(2\pi x) \cos(2\pi y)$ \\
$v(x, y, 0) = -\cos(2\pi x) \sin(2\pi y)$

\medskip
\noindent \textbf{Parameters:} \\
Set density $\rho = 1$ and kinematic viscosity $\nu = 1 \times 10^{-3}$. Simulate up to time $t = 1$. Choose a time step small enough to lead to a stable numerical solution considering the advection and viscous time scales.

\medskip
\noindent \textbf{Numerical outputs:} \\
Save the velocity field at times (e.g., $t = 0, 0.25, 0.5, 1.0$) in XDMF format.

\end{tcolorbox}

\paragraph{Solid Mechanics Problems}
\vspace{.25cm}

\begin{tcolorbox}[enhanced,breakable,colback=gray!10,colframe=gray!50,title={Solid Mechanics Problem 1 (Easy)}, label=sm_q1, 
    phantom={\phantomsection\label{box:sm_q1}}]

\noindent \textbf{Geometry:} \\
Let $\Omega = (0, 1.0) \times (0, 0.20)$~m be a rectangular plate.

\medskip
\noindent \textbf{Mesh:} \\
Uniform structured mesh with $20 \times 4$ subdivisions across $(x, y)$.

\medskip
\noindent \textbf{Model:} \\
Plane-stress linear elasticity for displacement $u = (u_x, u_y)$ in $\Omega$. \\
$\sigma$ is the Cauchy stress tensor and $n$ is the outward unit normal.

\medskip
\noindent \textbf{Material:} \\
Young's modulus $E = 200$~GPa, Poisson's ratio $\nu = 0.30$.

\medskip
\noindent \textbf{Boundary conditions:} \\
Left edge ($x = 0$): fixed, $u_x = 0, u_y = 0$. \\
Right edge ($x = 1$): prescribed displacement, $u_x = 0.001$~m, $u_y = 0$. \\
Top ($y = 0.20$) and bottom ($y = 0$): traction-free ($\sigma n = 0$).

\medskip
\noindent \textbf{Output:} \\
Save a color map of the horizontal displacement $u_x$ as \texttt{q1\_ux.png}. \\
Save the resulting displacement field in XDMF format.
\end{tcolorbox}

\begin{tcolorbox}[enhanced,breakable,colback=gray!10,colframe=gray!50,title={Solid Mechanics Problem 2 (Easy)}, label=sm_q2, 
    phantom={\phantomsection\label{box:sm_q2}}]

\noindent \textbf{Geometry:} \\
Let $\Omega = (0, 1.0) \times (0, 0.20)$~m be a rectangular plate.

\medskip
\noindent \textbf{Mesh:} \\
Use uniform structured mesh with $40 \times 8$ subdivisions across $(x, y)$.

\medskip
\noindent \textbf{Model:} \\
Plane-stress linear elasticity for displacement $u = (u_x, u_y)$ in $\Omega$. \\
$\sigma$ is the Cauchy stress tensor and $n$ is the unit outward normal.

\medskip
\noindent \textbf{Material:} \\
Young's modulus $E = 200$~GPa, Poisson's ratio $\nu = 0.30$.

\medskip
\noindent \textbf{Boundary conditions:} \\
Left edge ($x = 0$): fixed, $u_x = 0, u_y = 0$. \\
Top edge ($y = 0.20$): traction boundary condition is $t = \sigma n = (0, -2000)$~N/m. \\
Right edge ($x = 1.0$) and bottom edge ($y = 0$): traction-free, $\sigma n = 0$.

\medskip
\noindent \textbf{Output:} \\
Save a color map of the vertical displacement $u_y$ as \texttt{q2\_uy.png}. \\
Save the resulting displacement field in XDMF format.
\end{tcolorbox}

\begin{tcolorbox}[enhanced,breakable,colback=gray!10,colframe=gray!50,title={Solid Mechanics Problem 3 (Easy)}, label=sm_q3, 
    phantom={\phantomsection\label{box:sm_q3}}]

\noindent \textbf{Geometry:} \\
Let $\Omega = (0, 1.0) \times (0, 0.20)$~meter be a rectangular plate with a centered circular hole of radius $a = 0.05$~m at $(0.50, 0.10)$. 

\medskip
\noindent \textbf{Model:} \\
Plane-stress linear elasticity for displacement $u = (u_x, u_y)$ in $\Omega$. \\
$\sigma$ is the Cauchy stress tensor and $n$ is the outward unit normal.

\medskip
\noindent \textbf{Material:} \\
Young's modulus $E = 200$~GPa, Poisson's ratio $\nu = 0.30$.

\medskip
\noindent \textbf{Boundary conditions:} \\
Left edge ($x = 0$): fixed, $u_x = 0, u_y = 0$. \\
Right edge ($x = 1.0$): $\sigma n = (2 \text{ MPa} \cdot \text{m}, 0)$. \\
All other external boundaries (including the hole boundary): traction-free, $\sigma n = 0$.

\medskip
\noindent \textbf{Output:} \\
Compute the von Mises equivalent stress field (plane-stress). Save a color map as \texttt{q3\_vm.png}. \\
Save the resulting displacement field in XDMF format. \\
Report the maximum von Mises stress at the hole boundary and the stress concentration factor ($K_t = \sigma_{\max}/2 \text{ MPa}$).
\end{tcolorbox}

\begin{tcolorbox}[enhanced,breakable,colback=gray!10,colframe=gray!50,title={Solid Mechanics Problem 4 (Easy)}, label=sm_q4, 
    phantom={\phantomsection\label{box:sm_q4}}]
    
\noindent \textbf{Geometry:} \\
Let $\Omega = (0, 1.0) \times (0, 0.20)$~meter be a rectangular plate with two circular holes of radius $a = 0.04$~m centered at $(0.33, 0.10)$~m and $(0.67, 0.10)$~m. 

\medskip
\noindent \textbf{Model:} \\
Plane-stress linear elasticity for displacement $u = (u_x, u_y)$ in $\Omega$. \\
$\sigma$ the Cauchy stress tensor and $n$ is the unit outward normal.

\medskip
\noindent \textbf{Material:} \\
Young's modulus $E = 200$~GPa, Poisson's ratio $\nu = 0.30$.

\medskip
\noindent \textbf{Boundary conditions:} \\
Left edge ($x = 0$): clamped, $u_x = 0, u_y = 0$. \\
Right edge ($x = 1.0$): $\sigma n = (2 \text{ MPa} \cdot \text{m}, 0)$. \\
Remaining boundaries (top, bottom, and both hole boundaries): traction-free, $\sigma n = 0$.

\medskip
\noindent \textbf{Output:} \\
Compute the von Mises equivalent stress and save a color map as \texttt{q4\_vm.png}. \\
Save the resulting displacement field in XDMF format. \\
Report the maximum von Mises stress at the hole boundary and the stress concentration factor ($K_t = \sigma_{\max}/2 \text{ MPa}$).
\end{tcolorbox}

\begin{tcolorbox}[enhanced,breakable,colback=gray!10,colframe=gray!50,title={Solid Mechanics Problem 5 (Easy)}, label=sm_q5, 
    phantom={\phantomsection\label{box:sm_q5}}]

\noindent \textbf{Geometry:} \\
Let $\Omega = (0, 1.0) \times (0, 0.20)$~meter be a rectangular plate with a rectangular notch cut from the left edge defined by the region $(0, 0.06) \times (0.08, 0.12)$. 

\medskip
\noindent \textbf{Model:} \\
Plane-stress linear elasticity for displacement $u = (u_x, u_y)$ in $\Omega$. \\
$\sigma$ is the Cauchy stress tensor and $n$ is the outward unit normal.

\medskip
\noindent \textbf{Material:} \\
Young's modulus $E = 200$~GPa, Poisson's ratio $\nu = 0.30$.

\medskip
\noindent \textbf{Boundary conditions and loads:} \\
Left outer boundary ($x = 0$): fixed, $u_x = 0, u_y = 0$. \\
Right edge ($x = 1.0$): $\sigma n = (2 \text{ MPa} \cdot \text{m}, 0)$. \\
Top and bottom edges, and notch boundaries: traction-free, $\sigma n = 0$.

\medskip
\noindent \textbf{Output:} \\
Compute the von Mises equivalent stress and save a color map as \texttt{q5\_vm.png}. \\
Save the resulting displacement field in XDMF format.
\end{tcolorbox}

\begin{tcolorbox}[enhanced,breakable,colback=gray!10,colframe=gray!50,title={Solid Mechanics Problem 6 (Easy)}, label=sm_q6, 
    phantom={\phantomsection\label{box:sm_q6}}]

\noindent \textbf{Geometry:} \\
Let $\Omega = (0, 1.0) \times (0, 0.20)$~meter be a rectangular plate with a semicircular notch of radius $a = 0.05$~m removed from the top edge. The center of this semicircular notch is $[0.5, 0.2]$. 

\medskip
\noindent \textbf{Model:} \\
Plane-stress linear elasticity for displacement $u = (u_x, u_y)$ in $\Omega$. \\
$\sigma$ is the Cauchy stress tensor and $n$ the outward unit normal.

\medskip
\noindent \textbf{Material:} \\
Young's modulus $E = 200$~GPa, Poisson's ratio $\nu = 0.30$.

\medskip
\noindent \textbf{Boundary conditions:} \\
Bottom edge ($y = 0$): fixed, $u_x = 0, u_y = 0$. \\
Top edge ($y = 0.20$, on $x$ from $[0, 0.45) \cup (0.55, 1.0]$): $\sigma n = (0, -10 \text{ MPa} \cdot \text{m})$. \\
Both vertical sides ($x = 0$ and $x = 1$) and the notch arc: Traction-free, $\sigma n = (0, 0)$.

\medskip
\noindent \textbf{Output:} \\
Compute the von Mises equivalent stress (plane-stress) and save a color map as \texttt{q6\_vm.png}. \\
Save the resulting displacement field in XDMF format.
\end{tcolorbox}

\begin{tcolorbox}[enhanced,breakable,colback=gray!10,colframe=gray!50,title={Solid Mechanics Problem 7 (Easy)}, label=sm_q7, 
    phantom={\phantomsection\label{box:sm_q7}}]

\noindent \textbf{Geometry:} \\
Let $\Omega = (0, 1.0) \times (0, 0.20)$~meter be a rectangular plate in 2D plane-stress. \\
Define two material subdomains: \\
Top half $\Omega_{\text{Al}} = (0, 1.0) \times (0.10, 0.20)$~m: aluminum. \\
Bottom half $\Omega_{\text{Steel}} = (0, 1.0) \times (0.00, 0.10)$~m: steel.

\medskip
\noindent \textbf{Mesh:} \\
Use structured mesh $80 \times 16$ over $\Omega$.

\medskip
\noindent \textbf{Model:} \\
Small-strain, linear elasticity for displacement $u = (u_x, u_y)$ in $\Omega$. \\
$\sigma$ the Cauchy stress tensor and $n$ the unit outward normal. \\
There is a perfect bond at the material interface ($y = 0.10$~m). Enforce the continuity of displacement and traction across the interface.

\medskip
\noindent \textbf{Material:} \\
$\Omega_{\text{Al}}$: Young's modulus $E = 70$~GPa, Poisson's ratio $\nu = 0.30$. \\
$\Omega_{\text{Steel}}$: Young's modulus $E = 200$~GPa, Poisson's ratio $\nu = 0.30$.

\medskip
\noindent \textbf{Boundary conditions and loads:} \\
Left edge ($x = 0$): fixed, $u_x = 0, u_y = 0$. \\
Right edge ($x = 1$): uniform downward line traction $\sigma n = (0, -5000)$~N~m$^{-1}$. \\
Top ($y = 0.20$) and bottom ($y = 0$): traction-free, $\sigma n = 0$.

\medskip
\noindent \textbf{Output:} \\
Save a color map of displacement magnitude $|u|$ as \texttt{q7\_disp.png}. \\
Save the resulting displacement field in XDMF format.
\end{tcolorbox}

\begin{tcolorbox}[enhanced,breakable,colback=gray!10,colframe=gray!50,title={Solid Mechanics Problem 8 (Medium)}, label=sm_q8, 
    phantom={\phantomsection\label{box:sm_q8}}]

\noindent \textbf{Geometry:} \\
Let $\Omega = (0, 1.0) \times (0, 0.20)$ meter be a rectangular plate modelled in 2D plane-stress. \\
Use structured mesh with $50 \times 25$ subdivisions over $\Omega$.

\medskip
\noindent \textbf{Model:} \\
Small-strain, linear elasticity for displacement $u = (u_x, u_y)$ in $\Omega$. \\
$\sigma$ is the Cauchy stress tensor and $n$ is the unit outward normal.

\medskip
\noindent \textbf{Material (orthotropic, plane-stress):} \\
The material’s principal axes (1--2) are rotated by $\theta = 30$ degrees anticlockwise with respect to the geometrical $x$--$y$ axes.  \\
Orthotropic lamina properties in the local 1--2 axes: \\
$E_1 = 40$~GPa (Young's modulus along the principal 'grain' direction) \\
$E_2 = 10$~GPa (Young's modulus across the 'grain') \\
$G_{12} = 5$~GPa (Shear modulus in the 1--2 plane) \\
$\nu_{12} = 0.25$ (Poisson's ratio, for strain in direction 2 from a load in direction 1) \\
Use the standard plane-stress reduced stiffness $[Q]$ and rotate it to the global $x$--$y$ frame via the angle-transformed stiffness matrix at $\theta = 30$ degrees.

\medskip
\noindent \textbf{Boundary conditions:} \\
Bottom edge ($y = 0$): Fixed, $u_x = 0, u_y = 0$. \\
Top edge ($y = 0.20$): Uniform downward traction $\sigma n = (0, -10 \text{ MPa} \cdot \text{m})$. \\
Vertical sides ($x = 0$ and $x = 1$): Traction-free, $\sigma n = (0, 0)$.

\medskip
\noindent \textbf{Output:} \\
Save a color map of the horizontal displacement $u_x$ as \texttt{q8\_ux.png}. Also, save a color map of the von Mises stress as \texttt{q8\_vm.png}. \\
Also write fields to XDMF (\texttt{q8\_solution.xdmf}). \\
Also, save the displacement ($u$) and stress ($\sigma$) fields to \texttt{q8\_solution.xdmf}.
\end{tcolorbox}

\begin{tcolorbox}[enhanced,breakable,colback=gray!10,colframe=gray!50,title={Solid Mechanics Problem 9 (Medium)}, label=sm_q9, 
    phantom={\phantomsection\label{box:sm_q9}}]

\noindent \textbf{Geometry:} \\
Let $\Omega = (0, 1.0) \times (0, 0.20)$ meter be a rectangular plate modelled in 2D plane-stress. \\
Use structured mesh with $100 \times 20$ subdivisions over $\Omega$.

\medskip
\noindent \textbf{Model:} \\
Solve the linear elasticity equation for displacement $u = (u_x, u_y)$ in $\Omega$. \\
$\sigma$ is the Cauchy stress tensor and $n$ the unit outward normal.

\medskip
\noindent \textbf{Material:} \\
Poisson's ratio $\nu = 0.30$. \\
Young's modulus varies with height: $E(y) = 100 \text{ GPa} + 100 \text{ GPa} \times (y / 0.20)$, for $y \in [0, 0.20]$~m.

\medskip
\noindent \textbf{Boundary conditions and loads:} \\
Left edge ($x = 0$): fixed, $u_x = 0, u_y = 0$. \\
Right edge ($x = 1$): uniform traction $\sigma n = (2\times10^6 \text{ Pa} \cdot \text{m}, 0)$ per unit thickness. \\
Top ($y = 0.20$) and bottom ($y = 0$): traction-free, $\sigma n = 0$.

\medskip
\noindent \textbf{Output:} \\
Save a color map of displacement magnitude $|u|$ as \texttt{q9\_disp.png}. \\
Save the resulting displacement field in XDMF format.
\end{tcolorbox}

\begin{tcolorbox}[enhanced,breakable,colback=gray!10,colframe=gray!50,title={Solid Mechanics Problem 10 (Medium)}, label=sm_q10, 
    phantom={\phantomsection\label{box:sm_q10}}]

\noindent \textbf{Geometry:} \\
Let $\Omega = (0, 1.0) \times (0, 0.20)$~m be a rectangular strip analyzed in 2D.

\medskip
\noindent \textbf{Mesh:} \\
Use structured mesh $100 \times 20$ over $\Omega$.

\medskip
\noindent \textbf{Model:} \\
Small-strain linear elasticity for displacement $u = (u_x, u_y)$ on $\Omega$. Assume plane strain. \\
$\sigma$: Cauchy stress; $n$: outward unit normal.

\medskip
\noindent \textbf{Material:} \\
Young's modulus $E = 5$~MPa, Poisson's ratio $\nu = 0.49$.

\medskip
\noindent \textbf{Boundary conditions and loading:} \\
Left edge ($x = 0$): fixed, $u_x = 0, u_y = 0$. \\
Right edge ($x = 1.0$): prescribed displacement $u_x = 0.03$~m, $u_y = 0$. \\
Top ($y = 0.20$) and bottom ($y = 0$): traction-free, $\sigma n = 0$.

\medskip
\noindent \textbf{Output:} \\
Save a color map of displacement magnitude $|u|$ as \texttt{q10\_disp.png}. \\
Save the resulting displacement field in XDMF format.

\medskip
\noindent \textbf{Note:} \\
Because $\nu \approx 0.5$, pure displacement elements over-stiffen (volumetric locking). A mixed displacement-pressure formulation is standard practice for nearly incompressible solids.
\end{tcolorbox}

\begin{tcolorbox}[enhanced,breakable,colback=gray!10,colframe=gray!50,title={Solid Mechanics Problem 11 (Medium)}, label=sm_q11, 
    phantom={\phantomsection\label{box:sm_q11}}]

\noindent \textbf{Geometry:} \\
Let $\Omega = (0, 1.0) \times (0, 0.20)$~m be a rectangular strip with a circular hole of radius $a = 0.04$~m, centered at $(0.50, 0.10)$. 

\medskip
\noindent \textbf{Model:} \\
Small-strain linear elasticity for displacement $u = (u_x, u_y)$ on $\Omega$. \\
$\sigma$ is the Cauchy stress tensor and $n$ is the outward unit normal.

\medskip
\noindent \textbf{Material (plane strain, nearly incompressible):} \\
Young's modulus $E = 5$~MPa, Poisson's ratio $\nu = 0.49$.

\medskip
\noindent \textbf{Boundary conditions and loading:} \\
Left edge ($x = 0$): fixed, $u_x = 0, u_y = 0$. \\
Right edge ($x = 1.0$): prescribed displacement, $u_x = 0.001$~m, $u_y = 0$. \\
Top ($y = 0.20$) and bottom ($y = 0$), and the circular hole boundary: traction-free $\sigma n = 0$.

\medskip
\noindent \textbf{Output:} \\
Compute the von Mises equivalent stress and save a color map as \texttt{q11\_vm.png}. \\
Save a color map of the horizontal displacement $u_x$ as \texttt{q11\_ux.png}. \\
Save the resulting displacement field in XDMF format.

\medskip
\noindent \textbf{Note:} \\
Because $\nu \approx 0.5$, pure displacement elements over-stiffen (volumetric locking). A mixed displacement-pressure formulation is standard practice for nearly incompressible solids.
\end{tcolorbox}

\begin{tcolorbox}[enhanced,breakable,colback=gray!10,colframe=gray!50,title={Solid Mechanics Problem 12 (Medium)}, label=sm_q12, 
    phantom={\phantomsection\label{box:sm_q12}}]

\noindent \textbf{Geometry:} \\
Let $\Omega = (0, 1.0) \times (0, 0.20)$~m be a rectangular strip with a circular hole of radius $a = 0.04$~m, centered at $(0.50, 0.10)$. 

\medskip
\noindent \textbf{Model:} \\
The rectangular strip is modeled as large-deformation, isotropic incompressible neo-Hookean solid, where the unknown fields are displacement $u = (u_x, u_y)$ and hydrostatic pressure $p$. \\
$\sigma$ is the Cauchy stress tensor and $n$ is the outward unit normal.

\medskip
\noindent \textbf{Material (plane strain):} \\
Young's modulus $E = 5$~MPa and Poisson ratio $\nu = 0.5$. Use the standard incompressible neo-Hookean strain-energy with a pressure field $p$ to impose $J = 1$ (volume constraint) in a mixed ($u, p$) formulation.

\medskip
\noindent \textbf{Boundary conditions and loading:} \\
Left edge ($x = 0$): fixed, $u_x = 0, u_y = 0$. \\
Right edge ($x = 1.0$): prescribed displacement, $u_x = 0.060$~m, $u_y = 0$. \\
Top ($y = 0.20$), bottom ($y = 0$), and the circular hole boundary: traction-free, $\sigma n = 0$.

\medskip
\noindent \textbf{Output:} \\
Save a color map of hydrostatic pressure $p$ as \texttt{q12\_p.png}. \\
Compute and save a color map of von Mises stress (from the Cauchy stress) as \texttt{q12\_vm.png}. \\
Save the resulting displacement field in XDMF format.
\end{tcolorbox}

\begin{tcolorbox}[enhanced,breakable,colback=gray!10,colframe=gray!50,title={Solid Mechanics Problem 13 (Medium)}, label=sm_q13, 
    phantom={\phantomsection\label{box:sm_q13}}]

\noindent \textbf{Geometry:} \\
Let $\Omega = (0, 1.0) \times (0, 0.20)$~m be a rectangular strip with a circular hole of radius $a = 0.04$~m, centered at $(0.50, 0.10)$.

\medskip
\noindent \textbf{Model:} \\
The rectangular strip is modeled as large-deformation, isotropic incompressible Neo-Hookean solid, where the unknown fields are displacement $u = (u_x, u_y)$ and hydrostatic pressure $p$. \\
$\sigma$ the Cauchy stress tensor and $n$ is the outward unit normal in the deformed configuration.

\medskip
\noindent \textbf{Material (plane strain):} \\
Take $E = 5$~MPa, $\nu = 0.5$. Use the standard incompressible neo-Hookean strain-energy with a pressure field $p$ to impose $J = 1$ (volume constraint) in a mixed ($u, p$) formulation.

\medskip
\noindent \textbf{Boundary conditions and loading:} \\
Left edge ($x = 0$): fixed, $u_x = 0, u_y = 0$. \\
Hole boundary: uniform follower pressure $P_{\text{hole}} = 0.10$~MPa applied as a traction $\sigma n = -P_{\text{hole}} n$. \\
Right ($x = 1.0$), top ($y = 0.20$), and bottom ($y = 0$) edges: traction-free, $\sigma n = 0$.

\medskip
\noindent \textbf{Output:} \\
Save a magnified ($\times 5$) plot of the deformed configuration as \texttt{q13\_def.png}. \\
Compute the von Mises equivalent stress from the Cauchy stress and save a color map as \texttt{q13\_vm.png}. \\
Save the resulting displacement field in XDMF format.
\end{tcolorbox}

\begin{tcolorbox}[enhanced,breakable,colback=gray!10,colframe=gray!50,title={Solid Mechanics Problem 14 (Hard)}, label=sm_q14, 
    phantom={\phantomsection\label{box:sm_q14}}]

\noindent \textbf{Geometry:} \\
Let $\Omega = (0, 1.0) \times (0, 0.20)$~m be a rectangular strip with two circular holes of radius $a = 0.04$~m, centered at $(0.40, 0.10)$~m and $(0.60, 0.10)$~m. 

\medskip
\noindent \textbf{Model:} \\
The rectangular strip is modeled as large-deformation, isotropic quasi-incompressible Neo-Hookean solid with unknown displacement $u = (u_x, u_y)$ and hydrostatic pressure $p$. \\
$\sigma$ is the Cauchy stress tensor and $n$ is the outward unit normal to the boundary in the deformed configuration. The specimen is under plane strain conditions.

\medskip
\noindent \textbf{Material:} \\
Take $E = 5$~MPa and $\nu = 0.49$.

\medskip
\noindent \textbf{Boundary conditions and loading:} \\
Left edge ($x = 0$): fixed, $u_x = 0, u_y = 0$. \\
Right edge ($x = 1.0$): prescribed displacement, $u_x = +0.060$~m, $u_y = 0$. \\
Each hole boundary: uniform follower pressure $P_{\text{hole}} = 0.10$~MPa applied as a traction $\sigma n = -P_{\text{hole}} n$. \\
Top ($y = 0.20$) and bottom ($y = 0$): traction-free.

\medskip
\noindent \textbf{Output:} \\
Save a color map of the pressure field $p$ as \texttt{q14\_p.png}. \\
Compute and save a color map of the von Mises stress from the Cauchy stress as \texttt{q14\_vm.png}. \\
Save the resulting displacement field in XDMF format.
\end{tcolorbox}

\begin{tcolorbox}[enhanced,breakable,colback=gray!10,colframe=gray!50,title={Solid Mechanics Problem 15 (Hard)}, label=sm_q15, 
    phantom={\phantomsection\label{box:sm_q15}}]

\noindent \textbf{Geometry:} \\
Let $\Omega = (0, 1.20) \times (0, 0.20)$~m be a rectangular strip with three circular holes of radius $a = 0.03$~m, centered on the midline $y = 0.10$~m at $x = 0.30$~m, $0.60$~m, and $0.90$~m. 

\medskip
\noindent \textbf{Model:} \\
Finite-strain Saint-Venant-Kirchhoff elasticity. Assume plane strain. The unknown is the displacement field $u = (u_x, u_y)$. \\
Define: \\
$F = I + \nabla u$ (deformation gradient) \\
$E = 0.5 (F^T F - I)$ (Green-Lagrange strain) \\
$S = \lambda \text{tr}(E) I + 2 \mu E$ (second Piola-Kirchhoff stress) \\
$J = \det(F)$, and Cauchy stress $\sigma = (1/J) F S F^T$

\medskip
\noindent \textbf{Material:} \\
Use Lamé parameters: $\lambda = 5.769$~MPa, $\mu = 3.846$~MPa.

\medskip
\noindent \textbf{Boundary conditions and loading:} \\
Left edge ($x = 0$): fixed, $u_x = 0, u_y = 0$. \\
Right edge ($x = 1.20$): prescribed displacement, $u_x = +0.012$~m, $u_y = 0$. \\
Top ($y = 0.20$) and bottom ($y = 0$): traction-free. \\
Hole boundaries: traction-free.

\medskip
\noindent \textbf{Output:} \\
Save a plot of the deformed configuration as \texttt{q15\_def.png}. \\
Compute principal Green-Lagrange strains; save a color map of the maximum principal value $E_{\max}$ as \texttt{q15\_Emax.png}. \\
Define $s = S - (1/3) \text{tr}(S) I$ and $\sigma_{\text{vm}}(s) = \sqrt{1.5 (s:s)}$; save a color map as \texttt{q15\_vmS.png}. \\
Export the final displacement $u$ and $E_{\max}$ in XDMF format (\texttt{q15\_u.xdmf}, \texttt{q15\_Emax.xdmf}).

\medskip
\noindent \textbf{Notes:} \\
Use load stepping with Newton iterations. Use load stepping and stop when max principal Green-Lagrange strain $E_{\max} \le 0.03$; if $E_{\max}$ exceeds $0.03$, do not advance the load.
\end{tcolorbox}

\begin{tcolorbox}[enhanced,breakable,colback=gray!10,colframe=gray!50,title={Solid Mechanics Problem 16 (Hard)}, label=sm_q16, 
    phantom={\phantomsection\label{box:sm_q16}}]

\noindent \textbf{Geometry:} \\
Analyze a plate in plane strain using the symmetric quarter domain $[0, 100] \times [0, 180]$~mm with a quarter hole of radius $50$~mm at the origin. 

\medskip
\noindent \textbf{Model:} \\
Use small-strain von Mises elastoplasticity with associative flow and perfect plasticity. \\
Let $u = (u_x, u_y)$, $\varepsilon = 1/2 (\nabla u + \nabla u^T)$. \\
The strain tensor is split into two elastic and plastic strain as $\varepsilon_{ij} = \varepsilon^e_{ij} + \varepsilon^p_{ij}$, where $\varepsilon^e$ represents the elastic part and $\varepsilon^p$ the plastic part. The plastic part verifies $\varepsilon^p_{kk}=0$. Here, and in what follows, repeated indices indicate summation. \\
The quasi-static momentum balance is $\sigma_{ij,j} = 0$, where $\sigma_{ij} = (\lambda + 2/3 \mu) \varepsilon_{kk} \delta_{ij} + s_{ij}$.  \\
Here, $s_{ij} = 2 \mu ( e_{ij} - e^p_{ij} )$, where $e_{ij} = \varepsilon_{ij} - (1/3) \varepsilon_{kk} \delta_{ij}$ and $e^p_{ij}$ is the plastic deviatoric strain. \\
The plastic flow is given by $\Delta e^p_{ij} = \Delta \varepsilon^p  (3/2)  s_{ij} / q$, where $q = \sqrt{ (3/2) s_{ij} s_{ij} }$ is the equivalent stress, and $\Delta \varepsilon^p$ is the increment of equivalent plastic strain (the plastic multiplier), defined as $\Delta \varepsilon^p = \sqrt{2/3 \Delta \varepsilon^p_{ij} \Delta \varepsilon^p_{ij}}$. \\
The yield condition is $F = q - \sigma_Y \le 0$ and the plastic loading follow the Kuhn-Tucker conditions, i.e., $\Delta \varepsilon^p \ge 0$, $F \le 0$, $\Delta \varepsilon^p F = 0$. The initial plastic strain is zero.

\medskip
\noindent \textbf{Boundary conditions:} \\
Apply symmetry on $x = 0$ using $u_x = 0$ and zero traction in a direction tangent to $x=0$. \\
Apply symmetry on $y = 0$ using $u_y = 0$ and zero traction in a direction tangent to $y=0$. \\
At the top edge ($y = 180$), prescribe $u = (0, 1)$~mm. \\
Set the edge $x = 100$~mm and the hole boundary traction-free.

\medskip
\noindent \textbf{Parameters:} \\
Use $\lambda = 19.44$~GPa and $\mu = 29.17$~GPa. Set $\sigma_Y = 243$~MPa.

\medskip
\noindent \textbf{Output:} \\
Save the resulting displacement in an xdmf file.
\end{tcolorbox}

\paragraph{Multiphysics Problems}
\vspace{.25cm}

\begin{tcolorbox}[enhanced,breakable,colback=gray!10,colframe=gray!50,title={Multiphysics Problem 1 (Hard)}, label=mf_q1, 
    phantom={\phantomsection\label{box:mf_q1}}]

\noindent \textbf{Geometry:} \\
Solve the transport of a chemical inside an expanding circular disk $\Omega(t)$ with radius $R(t)$, where $R(t) = R_0 + s \cdot t$ with constant rate $s = 1.0 \times 10^{-4}$~m~s$^{-1}$. Here, $R_0$ is the initial radius and is equal to $0.05$~m. The boundary of this circular disk is denoted by $\Gamma(t)$.

\medskip
\noindent \textbf{Mesh:} \\
Unstructured triangular mesh of $\Omega(0)$ with characteristic size $h_0 \approx 1.0 \times 10^{-3}$~m. \\
The mesh motion is given by $w(x,t) = s x / \|x\|$ for $x \neq (0,0)$, $w(x,t) = 0$ for $x = 0$. 

\medskip
\noindent \textbf{Model:} \\
Solve the diffusion-reaction equation for concentration $c$ on a moving domain using an Arbitrary Lagrangian-Eulerian (ALE) description, with constant diffusivity $D$ and first-order decay equal to $\kappa c$, where $\kappa$ is the decay rate and is positive.

\medskip
\noindent \textbf{Boundary conditions:} \\
Zero total flux at the moving boundary: $(-D\nabla c - w c) \cdot n = 0$ on $\Gamma(t)$ for all $t \ge 0$, where $D$ is the diffusion coefficient. \\
Initial condition: $c(x,0) = 1$ for $x \in \Omega(0)$.

\medskip
\noindent \textbf{Parameters:} \\
Diffusivity: $D = 1.0 \times 10^{-5}$~m$^2$~s$^{-1}$. \\
Decay rate: $\kappa = 1.0 \times 10^{-4}$~s$^{-1}$. \\
Use the time step $\Delta t = 0.01$~s. Simulate up to 10 seconds.

\medskip
\noindent \textbf{Output:} \\
Save the output concentration as function of time in an XDMF file. \\
Report the total concentration in the domain after every 100 time steps.
\end{tcolorbox}

\begin{tcolorbox}[enhanced,breakable,colback=gray!10,colframe=gray!50,title={Multiphysics Problem 2 (Hard)}, label=mf_q2, 
    phantom={\phantomsection\label{box:mf_q2}}]

\noindent \textbf{Geometry:} \\
Unit square $\Omega = (0, 1) \times (0, 1)$.

\medskip
\noindent \textbf{Mesh:} \\
Use a structured mesh with $200 \times 200$ elements.

\medskip
\noindent \textbf{Model:} \\
Solve in $\Omega$ the Allen-Cahn curvature-flow equation for the phase field $\phi$, given by $\partial\phi/\partial t = -M ( (1/\varepsilon) W'(\phi) - \varepsilon \nabla^2\phi )$, where the double-well potential is $W(\phi) = 0.25 (\phi^2 - 1)^2$. Solve this equation using the finite element method. Use an implicit time stepping algorithm.

\medskip
\noindent \textbf{Boundary conditions:} \\
Impose homogeneous Neumann conditions on the entire boundary, namely $\nabla\phi \cdot n = 0$ on $\partial\Omega$. \\
Initialize the phase field using $\phi(x, y, 0) = \tanh(d_{\text{rect}}(x, y) / (\sqrt{2} \varepsilon))$, where $d_{\text{rect}}$ is a signed distance to the surface of a square that is centered at $(0.5, 0.5)$ with side length equal to $0.5$. The signed distance function is negative inside the square and positive outside. 

\medskip
\noindent \textbf{Parameters:} \\
Simulate up to the final time $T = 0.20$. Use an interface thickness $\varepsilon = 0.01$ and a mobility $M = 1.0$. Choose a time step $\Delta t = 1.0\text{e}-3$.

\medskip
\noindent \textbf{Output:} \\
Save the phase field, $\phi$, at each of the specified time steps: $t = 0.00, 0.05, 0.10$, and $0.20$. Store these files in XDMF format.
\end{tcolorbox}

\begin{tcolorbox}[enhanced,breakable,colback=gray!10,colframe=gray!50,title={Multiphysics Problem 3 (Hard)}, label=mf_q3, 
    phantom={\phantomsection\label{box:mf_q3}}]

\noindent \textbf{Geometry:} \\
Solve the spinodal decomposition of a chemical in the unit square $\Omega = (0, 1) \times (0, 1)$.

\medskip
\noindent \textbf{Model:} \\
Solve the non-dimensional Cahn-Hilliard equations in mixed form for the concentration $c$ and chemical potential $\mu$: \\
$\partial c/\partial t = \nabla \cdot ( M(c) \nabla \mu )$, with $\mu = 3\alpha \mu_c - \nabla^2 c$ and $\mu_c = (0.5/ \theta) \ln(c/(1 - c)) + 1 - 2c$. \\
Use the degenerate mobility $M(c) = c(1 - c)$. \\
Initialize with $c(x, y, 0) = \bar{c} + r(x, y)$, where $\bar{c} = 0.63$ and $r$ is a zero-mean uniform perturbation in $[-0.05, 0.05]$. \\
Advance in time using a backward Euler scheme.

\medskip
\noindent \textbf{Boundary conditions:} \\
Impose periodic boundary conditions for both $c$ and $\mu$ on $\partial\Omega$.

\medskip
\noindent \textbf{Parameters:} \\
Set $\theta = 1.5$, $\alpha = 3000$, and final time $T = 0.04$. \\
Report the fields at $t = 0, 3\text{e}-6, 1\text{e}-4, 1\text{e}-3$, and $4\text{e}-2$ in XDMF format.

\medskip
\noindent \textbf{Output:} \\
Save the concentration and chemical potential fields at $t = 0, 3\text{e}-6, 1\text{e}-4, 1\text{e}-3$, and $4\text{e}-2$ to a time-series file named \texttt{cahn\_hilliard.xdmf}

\medskip
\noindent \textbf{Hints:} \\
Use adaptive time stepping: start with $\Delta t$ in the range $1\text{e}-7$ to $5\text{e}-7$ and then increase or reduce the time step based on the nonlinear iterations required for convergence. Make sure the simulation can recover if the time step selected is too large. \\
Discretize by splitting the fourth-order equation into two coupled second-order equations and solve them with linear finite elements.
\end{tcolorbox}

\begin{tcolorbox}[enhanced,breakable,colback=gray!10,colframe=gray!50,title={Multiphysics Problem 4 (Hard)}, label=mf_q4, 
    phantom={\phantomsection\label{box:mf_q4}}]

\noindent \textbf{Geometry:} \\
Solve the coupled Stokes-Darcy problem on the rectangle $\Omega = (0, \pi) \times (-1, 1)$. The upper subdomain is the Stokes region $\Omega_S = (0, \pi) \times (0, 1)$ and the lower subdomain is the Darcy region $\Omega_D = (0, \pi) \times (-1, 0)$. The interface is $\Gamma = (0, \pi) \times \{0\}$ with unit normal $n = (0, -1)$ pointing from $\Omega_S$ into $\Omega_D$ and unit tangent $t = (1, 0)$. 

\medskip
\noindent \textbf{Model:} \\
In $\Omega_S$, the flow is incompressible Stokes with a body force: $\nabla \cdot u_S = 0$ and $-\nabla \cdot \sigma(u_S, p_S) = b$, with $\sigma = -p_S I + 2 \nu \text{sym}(\nabla u_S)$. Here, $\text{sym}(\nabla u_S)$ denotes the symmetric part of $\nabla u_S$. \\
In $\Omega_D$, the flow is incompressible Darcy: $\nabla \cdot u_D = 0$ and $u_D = -(k/\mu) \nabla p_D$. \\
On $\Gamma$, impose mass continuity $u_S \cdot n = u_D \cdot n$, normal traction balance $n \cdot \sigma n = -p_D/\rho$, and the Beavers-Joseph-Saffman tangential condition $(\alpha/\sqrt{k}) (u_S - u_D)\cdot t = - t \cdot \sigma n$. \\
Here, $u_S$ and $u_D$ are the fluid velocity in the Stokes and Darcy subdomains respectively, while $p_S$ and $p_D$ are the fluid pressure in the Stokes and Darcy subdomains, respectively.

\medskip
\noindent \textbf{Boundary conditions:} \\
Dirichlet boundary conditions for both Stokes and Darcy boundaries. \\
Stokes (on $\partial\Omega_S = \{y=1\} \cup \{x=0\}\times\{y\in[0,1]\} \cup \{x=\pi\}\times\{y\in[0,1]\}$): \\
$u_S(x,y) = [ w'(y) \cos x , w(y) \sin x ]$. \\
Darcy ($\partial\Omega_D = \{y=-1\} \cup \{x=0\}\times\{y\in[-1,0]\} \cup \{x=\pi\}\times\{y\in[-1,0]\}$): \\
$p_D(x,y) = \rho g \exp(y) \sin x$. \\
$w(y) = -K - (g y)/(2\nu) + (K/2 - \alpha g/(4\nu^2)) y^2$ and $K = k \rho g / \mu$.

\medskip
\noindent \textbf{Parameters:} \\
Take $g = 1, \rho = 1, \nu = 1, k = 1, K = 1, \alpha = 1$. \\
Body force $b$: \\
$b_x(x,y) = [ (\nu K - (\alpha g)/(2\nu)) y - g/2 ] \cos(x)$ \\
$b_y(x,y) = [ ( (\nu K)/2 - (\alpha g)/(4\nu) ) y^2 - (g/2) y + ( (\alpha g)/(2\nu) - 2\nu K ) ] \sin(x)$

\medskip
\noindent \textbf{Output:} \\
Save the Stokes velocity and Darcy pressure in XDMF format.
\end{tcolorbox}

\begin{tcolorbox}[enhanced,breakable,colback=gray!10,colframe=gray!50,title={Multiphysics Problem 5 (Hard)}, label=mf_q5, 
    phantom={\phantomsection\label{box:mf_q5}}]
Analyze the fluid flow in a 2D elastic tube using a fluid structure interaction model.\\

\noindent \textbf{Geometry:} \\
The fluid domain is a rectangle of length $6$~cm and height $1$~cm. The upper and lower walls are modeled as 2-D linear-elastic solids of uniform thickness $0.1$~cm, attached along the top and bottom boundaries of the fluid domain (so the outer faces of the upper and lower walls lie at $y = 1.1$~cm and $y = -0.1$~cm, respectively). 

\medskip
\noindent \textbf{Model:} \\
The fluid is incompressible and is solved in the moving domain using the Arbitrary Lagrangian Eulerian (ALE) form of the Navier-Stokes equations. The solid walls are small-strain, 2-D (plane-strain) linear-elastic bodies. \\
Enforce no-slip (fluid velocity equals wall velocity) and traction balance (fluid traction matches solid traction) at the fluid-solid interface.

\medskip
\noindent \textbf{Boundary conditions:} \\
At the inlet ($x = 0$), apply the normal-traction condition on the fluid: \\
$\sigma_f n_f = [ -(2\cdot10^4)/2 \cdot (1 - \cos(\pi t / 2.5\cdot10^{-3})), 0 ]^T$ for $t < 0.005$~s, and $\sigma_f n_f = 0$ thereafter. \\
At the outlet ($x = 6$~cm), impose zero traction, $\sigma_f n_f = 0$. \\
Here, $\sigma_f$ is the Cauchy stress tensor of the fluid and $n_f$ is the outward unit normal vector to the fluid domain. \\
The outer faces of the lower and upper walls are traction free. \\
The initial conditions are zero displacement and zero velocity for the solid and fluid domain respectively.

\medskip
\noindent \textbf{Parameters:} \\
Use a fluid viscosity $\mu_f = 0.003$~poise and a fluid density $\rho_f = 1$~g$\cdot$cm$^{-3}$. \\
Solid density $\rho_s = 1.1$~g$\cdot$cm$^{-3}$, and Poisson ratio $\nu_s = 0.49$, Young's modulus $E_s = 3.0\times10^5$~Pa. Use a mixed displacement-pressure formulation for the discretization of the solids mechanics equations. \\
The simulation time step is $\Delta t = 1.0\times10^{-4}$~s.

\medskip
\noindent \textbf{Numerical outputs:} \\
Save the output velocity and displacement at time $0.005$s and $0.1$s in XDMF format.
\end{tcolorbox}

\begin{tcolorbox}[enhanced,breakable,colback=gray!10,colframe=gray!50,title={Multiphysics Problem 6 (Hard)}, label=mf_q6, 
    phantom={\phantomsection\label{box:mf_q6}}]

Analyze the motion of a solid flag in a 2D flow channel using a fluid structure interaction model. \\

\noindent \textbf{Geometry:} \\
The fluid domain is a 2D channel of length $2.5$~m and height $0.41$~m. The flag is modeled as a rectangular elastic solid of length $0.35$~m and thickness $0.02$~m with the right-bottom corner placed at $(0.60 \text{ m}, 0.19 \text{ m})$ in the reference configuration. \\[.2cm]
The pole of the flag is modeled as circular disk of radius $0.05$~m centered at $(0.20 \text{ m}, 0.20 \text{ m})$. The pole is assumed to be rigid and we remove this disk from our computational domain. Since this disk intersects the fluid and solid domain, portions of both the solid and fluid domain are removed.  \\[.2cm]
The boundary of the circular disk is denoted by $\Gamma = \Gamma_s \cup \Gamma_f$, where $\Gamma_f$ and $\Gamma_s$ are the partitions of $\Gamma$ that are in contact with the fluid and solid domain, respectively.

\medskip
\noindent \textbf{Model:} \\
The fluid is incompressible and is solved in the moving domain using the Arbitrary Lagrangian Eulerian form of the Navier-Stokes equations. The solid flag is modeled using the St. Venant-Kirchhoff model. \\[.2cm]
Enforce no-slip (fluid velocity equals solid velocity) and traction balance (fluid traction equals solid traction) at the fluid-solid interface.

\medskip
\noindent \textbf{Boundary conditions:} \\
At the inlet ($x = 0$), prescribe a parabolic velocity profile with: \\
$u(t, 0, y) = \begin{cases} u_y(0,y)  \frac{1 - \cos(\pi t / 2)}{2}, & \text{if } t < 2.0 \text{ sec.} \\ u_y(0,y), & \text{if } t \ge 2.0 \text{ sec.} \end{cases}$ \\
where $u_y(0,y) = 1.5 \bar{U}  y (H - y) / (H/2)^2$ with $H = 0.41$ and $\bar{U} = 1$~m/s. \\
At the outlet ($x = 2.5$~m), use a traction free boundary condition. \\
Impose no-slip and no penetration on the top and bottom channel walls. \\
At $\Gamma_f$: Impose no slip and no penetration for the fluid. \\
At $\Gamma_s$: Impose zero displacement for the solid flag.

\medskip
\noindent \textbf{Parameters:} \\
Use fluid density $\rho_f = 1000$~kg/m$^3$ and kinematic viscosity $\nu_f = 1.0\times10^{-3}$~m$^2$/s. Use solid density $\rho_s = 10000$~kg/m$^3$, Poisson ratio $\nu_s = 0.4$, shear modulus $\mu_s = 0.5\times10^6$~Pa.

\medskip
\noindent \textbf{Numerical outputs:} \\
Save the fluid velocity and pressure fields and the beam displacement in XDMF format. \\
Report the displacement components of point A with time, where the reference configuration of point A is given by $A(t=0) = (0.60 \text{ m}, 0.20 \text{ m})$.
\end{tcolorbox}

\begin{tcolorbox}[enhanced,breakable,colback=gray!10,colframe=gray!50,title={Multiphysics Problem 7 (Hard)}, label=mf_q7, 
    phantom={\phantomsection\label{box:mf_q7}}]

\noindent \textbf{Geometry:} \\
Let $\Omega = [0, 1.0] \times [0, 0.20]$~m be a 2D rectangular channel. \\
The domain $\Omega$ is divided into two subdomains: \\
A porous filter: $\Pi = [0.4, 0.6] \times [0, 0.20]$~m.  \\
A free-fluid region: $\Omega_f = \Omega \setminus \Pi$. \\
The interface between them is $\partial\Pi$.

\medskip
\noindent \textbf{Model:} \\
Solve a coupled, steady, incompressible flow problem for velocity $u = (u_x, u_y)$ and pressure $p$. The governing equations differ by subdomain: \\
In $\Omega_f$ (Fluid): Steady Navier-Stokes equations. \\
In $\Pi$ (Porous): Steady incompressible Brinkman equations given by \\
$-\nabla p + \mu \nabla^2 u - (\mu/K) u = 0$ \\
$\nabla \cdot u = 0$ \\
At $\partial\Pi$ (Interface): Continuity of velocity and traction is enforced.

\medskip
\noindent \textbf{Boundary Conditions:} \\
Inlet ($x = 0$): $u_x(y) = 6 \bar{U} y (H - y) / H^2$, with $\bar{U} = 1.0$~m~s$^{-1}$. $u_y = 0$. \\
Walls ($y = 0$ and $y = H$): No-slip and no penetration. \\
Outlet ($x = 1.0$): traction-free, $(-pI + \mu(\nabla u + \nabla u^T)) n = 0$.

\medskip
\noindent \textbf{Parameters:} \\
Density ($\rho$): $1.0$~kg~m$^{-3}$ \\
Dynamic viscosity ($\mu$): $0.01$~Pa$\cdot$s. \\
Permeability ($K$): $1.0 \times 10^{-6}$~m$^2$.

\medskip
\noindent \textbf{Output:} \\
Save a color map of the velocity magnitude $|u|$ as \texttt{q14\_speed.png}. \\
Compute the pressure drop ($\Delta p$) across the porous block by sampling $p$ at the centerline just before and just after $\Pi$. Save this value to \texttt{q14\_dp.txt}. \\
Also, save the velocity field ($u$) and the pressure field ($p$) for the entire domain to \texttt{q14\_solution.xdmf}.
\end{tcolorbox}

\begin{tcolorbox}[enhanced,breakable,colback=gray!10,colframe=gray!50,title={Multiphysics Problem 8 (Hard)}, label=mf_q8, 
    phantom={\phantomsection\label{box:mf_q8}}]

\noindent \textbf{Geometry:} \\
The domain $[0, 1] \times [0, H]$~m is split into two adjacent rectangular subdomains: \\
Free Fluid ($\Omega_f$): $[0, 0.6] \times [0, H]$~m \\
Porous Medium ($\Omega_p$): $[0.6, 1.0] \times [0, H]$~m \\
Interface ($\Gamma$): The boundary between them at $x = 0.6$. 

\medskip
\noindent \textbf{Model:} \\
This is a coupled problem solving for velocity and pressure in each subdomain. \\
In $\Omega_f$ (Fluid): Steady Stokes equations for velocity $u_f$ and pressure $p_f$. \\
In $\Omega_p$ (Porous): Steady Darcy flow for velocity $u_p$ and pressure $p_p$. \\
$u_p = -(K/\mu) \nabla p_p$ \\
$\nabla \cdot u_p = 0$

\medskip
\noindent \textbf{Interface Conditions (on $\Gamma$ at $x = 0.6$)} \\
The two models are coupled by three conditions: \\
Continuity of Normal Velocity: $u_f \cdot n = u_p \cdot n$, where $n$ is the normal vector. \\
Continuity of Pressure: $p_f = p_p$. \\
Impose no slip for the Stokes velocity at the interface: $u_t = 0$, where $u_t$ is the tangential velocity on the fluid side.

\medskip
\noindent \textbf{Boundary Conditions (External)} \\
Inlet ($x = 0$, on $\Omega_f$): Prescribed velocity: $u_x(y) = 6 \bar{U} y (H - y) / H^2$, $u_y = 0$. \\
Fluid Walls ($y = 0$ and $y = H$, on $\Omega_f$): No-slip and no penetration. \\
Porous Walls ($y = 0$ and $y = H$, on $\Omega_p$): No-flux (impermeable) \\
Outlet ($x = 1.0$, on $\Omega_p$): Fixed pressure, $p_p = 0$ (gauge pressure).

\medskip
\noindent \textbf{Parameters:} \\
Dynamic Viscosity ($\mu$): $0.02$~Pa$\cdot$s \\
Permeability ($K$): $1.0 \times 10^{-6}$~m$^2$ \\
Mean Inlet Speed ($\bar{U}$): $0.1$~m~s$^{-1}$ \\
Channel Height ($H$): $0.2$~m

\medskip
\noindent \textbf{Output:} \\
Save the interface profiles for normal velocity ($u_x$) and tangential velocity ($u_y$) along $\Gamma$ (from $y=0$ to $y=H$) to \texttt{q15\_interface.csv}. \\
Save a color map of the pressure field $p$ (for both $p_f$ and $p_p$) as \texttt{q15\_p.png}. \\
Also, save the combined velocity field ($u$) and pressure field ($p$) for the entire domain to \texttt{q15\_solution.xdmf}.
\end{tcolorbox}

\end{appendices}

\bibliography{sample}

\end{document}